\documentclass[11pt,a4paper]{article}
\usepackage{jcappub} 
\usepackage{xspace}
\usepackage{hyperref}
\usepackage[noabbrev]{cleveref}
\usepackage{orcidlink}
\usepackage{amsmath}
\newcommand{\lya}{Ly$\alpha$\xspace}
\newcommand{\lyaxlya}{Ly$\alpha\times$Ly$\alpha$}
\newcommand{\lyaxqso}{Ly$\alpha\times$QSO}

\newcommand{\hMpc}{h^{-1}\,\mathrm{Mpc}}
\newcommand{\rp}{r_\parallel}
\newcommand{\rt}{r_\perp}
\newcommand{\zeff}{z_{\rm eff}}

\defcitealias{DESI.DR2.BAO.lya}{\texttt{DESIDR2-Ly$\alpha$}}
\newcommand{\DESIDRIILya}{\citetalias{DESI.DR2.BAO.lya}}

\defcitealias{DESI.DR2.BAO.cosmo}{\texttt{DESIDR2-Cosmo}}
\newcommand{\DESIDRIICosmo}{\citetalias{DESI.DR2.BAO.cosmo}}

\title{\boldmath Probing the matter-dominated expansion with multi-redshift Lyman-$\alpha$ BAO from DESI DR2}

\affiliation{Affiliations are in Appendix \ref{sec:affiliations}}
\emailAdd{herreraa@iap.fr}

\author[a,b]{{Hiram~K.~Herrera-Alcantar}\orcidlink{0000-0002-9136-9609},}
\author[c]{{Julien~Guy}\orcidlink{0000-0001-9822-6793},}
\author[d]{{Alma~X.~Gonzalez-Morales}\orcidlink{0000-0003-4089-6924},}
\author[b]{{Eric~Armengaud}\orcidlink{0000-0001-7600-5148},}
\author[d]{{Edwin~L.~Pérez-Ochoa}\orcidlink{0000-0002-9021-4381},}
\author[e,f]{{Cristhian~Garcia-Quintero}\orcidlink{0000-0003-1481-4294},}
\author[c]{{J.~Aguilar},}
\author[g]{{S.~Ahlen}\orcidlink{0000-0001-6098-7247},}
\author[h]{{F.~Beutler}\orcidlink{0000-0003-0467-5438},}
\author[i,j]{{D.~Bianchi}\orcidlink{0000-0001-9712-0006},}
\author[c]{{A.~Brodzeller}\orcidlink{0000-0002-8934-0954},}
\author[k]{{D.~Brooks},}
\author[c]{{E.~Chaussidon}\orcidlink{0000-0001-8996-4874},}
\author[c]{{T.~Claybaugh}\orcidlink{0000-0002-5024-6987},}
\author[c]{{A.~Cuceu}\orcidlink{0000-0002-2169-0595},}
\author[l]{{K.~S.~Dawson}\orcidlink{0000-0002-0553-3805},}
\author[m]{{A.~de la Macorra}\orcidlink{0000-0002-1769-1640},}
\author[n]{{Arjun~Dey}\orcidlink{0000-0002-4928-4003},}
\author[c,o]{{S.~Ferraro}\orcidlink{0000-0003-4992-7854},}
\author[p,q]{{A.~Font-Ribera}\orcidlink{0000-0002-3033-7312},}
\author[r,s]{{J.~E.~Forero-Romero}\orcidlink{0000-0002-2890-3725},}
\author[t,u,v]{{E.~Gaztañaga}\orcidlink{0000-0001-9632-0815},}
\author[w]{{G.~Gutierrez}\orcidlink{0000-0003-0825-0517},}
\author[x]{{C.~Hahn}\orcidlink{0000-0003-1197-0902},}
\author[y,z,aa]{{K.~Honscheid}\orcidlink{0000-0002-6550-2023},}
\author[ab,ac]{{D.~Huterer}\orcidlink{0000-0001-6558-0112},}
\author[ad]{{M.~Ishak}\orcidlink{0000-0002-6024-466X},}
\author[e,ae]{{T.~Karim}\orcidlink{0000-0002-5652-8870},}
\author[af]{{R.~Kehoe}\orcidlink{0000-0002-7101-697X},}
\author[ag]{{D.~Kirkby}\orcidlink{0000-0002-8828-5463},}
\author[c]{{A.~Kremin}\orcidlink{0000-0001-6356-7424},}
\author[k]{{O.~Lahav}\orcidlink{0000-0002-1134-9035},}
\author[c]{{A.~Lambert}\orcidlink{0009-0003-5658-2601},}
\author[c]{{M.~Landriau}\orcidlink{0000-0003-1838-8528},}
\author[ah]{{L.~Le~Guillou}\orcidlink{0000-0001-7178-8868},}
\author[ai,q]{{M.~Manera}\orcidlink{0000-0003-4962-8934},}
\author[y,aj,aa]{{P.~Martini}\orcidlink{0000-0002-4279-4182},}
\author[n]{{A.~Meisner}\orcidlink{0000-0002-1125-7384},}
\author[p,q]{{R.~Miquel}\orcidlink{0000-0002-6610-4836},}
\author[ak]{{J.~Moustakas}\orcidlink{0000-0002-2733-4559},}
\author[m]{{A.~Muñoz-Gutiérrez}\orcidlink{0000-0002-3565-0148},}
\author[u]{{S.~Nadathur}\orcidlink{0000-0001-9070-3102},}
\author[d,al]{{G.~Niz}\orcidlink{0000-0002-1544-8946},}
\author[am,an]{{E.~Paillas}\orcidlink{0000-0002-4637-2868},}
\author[b,c]{{N.~Palanque-Delabrouille}\orcidlink{0000-0003-3188-784X},}
\author[ao,ap,aq]{{W.~J.~Percival}\orcidlink{0000-0002-0644-5727},}
\author[c,ar,o]{{C.~Poppett}\orcidlink{0000-0003-0512-5489},}
\author[as]{{F.~Prada}\orcidlink{0000-0001-7145-8674},}
\author[at]{{I.~P\'erez-R\`afols}\orcidlink{0000-0001-6979-0125},}
\author[au]{{C.~Ravoux}\orcidlink{0000-0002-3500-6635},}
\author[av]{{G.~Rossi},}
\author[aw]{{R.~Ruggeri}\orcidlink{0000-0002-0394-0896},}
\author[ax,ay]{{L.~Samushia}\orcidlink{0000-0002-1609-5687},}
\author[az]{{E.~Sanchez}\orcidlink{0000-0002-9646-8198},}
\author[ba]{{C.~Saulder}\orcidlink{0000-0002-0408-5633},}
\author[c]{{D.~Schlegel}\orcidlink{0000-0002-5042-5088},}
\author[ab,ac]{{M.~Schubnell},}
\author[bb]{{H.~Seo}\orcidlink{0000-0002-6588-3508},}
\author[c]{{J.~Silber}\orcidlink{0000-0002-3461-0320},}
\author[ac]{{G.~Tarl\'{e}}\orcidlink{0000-0003-1704-0781},}
\author[n]{{B.~A.~Weaver},}
\author[b]{{C.~Yèche}\orcidlink{0000-0001-5146-8533},}
\author[c]{{R.~Zhou}\orcidlink{0000-0001-5381-4372},}

\usepackage{arydshln}
\usepackage{enumitem}

\date{\today}
\abstract{
We present a multi-redshift Baryon Acoustic Oscillations (BAO) analysis of the DESI Data Release 2 (DR2) Lyman-$\alpha$ (Ly$\alpha$) forest, splitting the forest auto-correlation and its cross-correlation with quasars into three redshift bins. We obtain BAO measurements at effective redshifts $z_{\rm eff} = 2.13$, $2.40$, and $2.81$ with $\sim2.0$--$2.5\%$ precision per bin in the radial and transverse directions, corresponding to $\sim1.1$--$1.2\%$ precision for the isotropic BAO measurement. Using the same data products and modeling framework as the DESI DR2 Ly$\alpha$ BAO analysis, we validate the pipeline on $400$ synthetic datasets and find unbiased BAO recovery with well-calibrated uncertainties. The measurements show an increase in the isotropic dilation parameter $D_V/r_d$ from $30.26\pm0.39$ to $32.22\pm0.47$ and in the Alcock-Paczyński parameter $D_M/D_H$ from $3.96\pm0.15$ to $5.63^{+0.22}_{-0.24}$. The Hubble distance $D_H/r_d$ decreases from $9.40\pm0.20$ to $7.22\pm0.17$, providing a direct measurement of the expansion history consistent with $\Lambda$CDM and the expected matter-dominated scaling, with $H(z)\propto(1+z)^n$ giving $n=1.34\pm0.16$. The redshift split also provides a self-consistent measurement of clustering evolution: the Ly$\alpha$ forest bias evolves as $(1+z)^\gamma$ with $\gamma_\alpha=3.05\pm0.16$, the RSD parameter has a redshift evolution described by $\gamma_\beta=-0.97\pm0.26$, and the quasar bias evolves with $\gamma_Q=1.56\pm0.23$, consistent with independent quasar clustering measurements. Combining these three-bin BAO measurements with DESI DR2 galaxy and quasar BAO measurements yields cosmological constraints consistent with the single-bin Ly$\alpha$ BAO analysis in flat $\Lambda$CDM and $w_0w_a$CDM and improves curvature constraints by $\sim12\%$ in $\Lambda$CDM$+\Omega_\mathrm{K}$.
}

\begin{document}
\maketitle
\flushbottom
\section{Introduction}\label{sec:introduction}

The Dark Energy Spectroscopic Instrument \citep[DESI;][]{DESI2016b.Instr, DESI2022.KP1.Instr} is a highly multiplexed spectroscopic survey operating on the Mayall 4-meter telescope at Kitt Peak National Observatory, with a wide-field corrector~\citep{Corrector.Miller.2023} and 5000 fiber-positioning robots~\citep{FiberSystem.Poppett.2024} that together cover a $\sim3.2^\circ$ field. Having completed its originally planned five-year survey~\citep{SurveyOps.Schlafly.2023} over roughly $14{,}000\,\mathrm{deg}^2$, DESI is now continuing into an eight-year campaign covering about $17{,}000\,\mathrm{deg}^2$, and has so far observed more than 40 million galaxies and quasars.

The DESI collaboration has been reporting precise measurements of 
the transverse comoving distance $D_M$ and the Hubble distance $D_H$, both relative to the sound horizon $r_d$, across seven tracer redshift bins. Using its first Data Release~\citep[DR1;][]{DESI2024.I.DR1}, DESI carried out its first cosmological analysis~\citep{DESI2024.VI.KP7A}, which used over 6 million extragalactic objects in the redshift range $0.1 < z < 4.2$ to measure the baryon acoustic oscillation (BAO) scale, with spectra processed by the DESI pipeline \citep{Spectro.Pipeline.Guy.2023}. These measurements included the auto-correlation of more than 420,000 Lyman-$\alpha$ (\lya) forest spectra and their cross-correlation with more than 700,000 quasars, at an effective redshift of $\zeff = 2.33$ \citep{DESI2024.IV.KP6}.

With the second data release (DR2), the \lya forest sample grew to over 820,000 quasar spectra and more than 1.2 million quasar positions for the cross-correlation~\citep{DESI.DR2.BAO.lya}, roughly doubling the DR1 sample. The DR2 \lya BAO analysis reports $D_H(\zeff)/r_d = 8.632 \pm 0.098\,{\rm (stat)} \pm 0.026\,{\rm (sys)}$ and $D_M(\zeff)/r_d = 38.99 \pm 0.52\,{\rm (stat)} \pm 0.12\,{\rm (sys)}$ at $\zeff = 2.33$, corresponding to a combined isotropic precision of $0.65\%$, the most precise large-scale-structure BAO measurement at $z > 2$ to date, and the first \lya BAO analysis precise enough to warrant a dedicated systematic uncertainty term. Combined with galaxy and quasar BAO tracers spanning $0.1 \leq z \leq 2.1$ \citep{DESI.DR2.BAO.cosmo}, these results show a mild $2.3\sigma$ tension with parameters inferred from the CMB alone, with the preference for a dynamical dark energy model over $\Lambda$CDM reaching $3\text{--}4\sigma$ when CMB and supernova data are included.

The \lya\ forest is the highest-redshift large-scale-structure tracer accessible to DESI, yet all previous DESI \lya\ BAO analyses have compressed its information into a single effective redshift. By dividing the sample into multiple redshift bins, one can instead track the evolution of the transverse and radial BAO distance scales across cosmic time while simultaneously measuring the redshift evolution of the \lya\ forest astrophysical parameters. A redshift split is also motivated by the strong evolution of the \lya\ forest bias across the redshift range probed by the sample. While the standard single-bin analysis averages over this evolution, a multi-redshift measurement provides a direct consistency check of the impact of this averaging on the recovered parameters.

First attempts at two-bin redshift splits were made for eBOSS DR14~\cite{deSainteAgathe:DR14,Blomqvist:DR14} and DR16~\citep{dmdB:DR16}. However, those analyses were statistics-limited, and were therefore not used to draw cosmological conclusions; nonetheless, they demonstrated the feasibility of the method and its potential to constrain the redshift evolution of astrophysical parameters beyond the BAO signal itself. The substantially larger quasar sample now available with DESI removes this limitation and enables a more thorough exploration of this multi-redshift approach, which may inform the design of future \lya\ BAO analyses.

In this context, distance measurements at high redshift provide a unique window into cosmology. At $z \gtrsim 2$, where dark energy is dynamically subdominant, they probe the expansion history of the matter-dominated era and, through measurements of $D_M/r_d$ and $D_H/r_d$, retain sensitivity to pre-recombination physics. They therefore provide constraints on extensions to the standard cosmological model, including spatial curvature, the effective number of relativistic species, neutrino properties, non-standard recombination histories, early dark energy scenarios, and non-standard dark matter models~\citep[see e.g.,][]{McDonald:2007,Xia:2009,Knox:2006curvature,Chen:2025curvature,Chudaykin:2018,Sailer:2021highz,Weiner:2026}.

In this work, we present BAO measurements from the \lya forest auto-correlation and the \lya-quasar cross-correlation using DESI DR2 data, analyzed in three distinct redshift bins at effective redshifts $\zeff = 2.13$, $2.40$, and $2.81$. Resolving the radial BAO scale $D_H(z)/r_d$ across these bins enables a direct, cosmology-independent test of the matter-dominated expansion history over $2 \lesssim z \lesssim 3$, a regime not directly accessible to lower-redshift galaxy surveys. We describe our methodology for constructing the redshift-binned correlations, fitting for the BAO peak position, measuring the evolution of the \lya forest astrophysical parameters, and combining the resulting distance measurements into cosmological constraints.

This paper is organized as follows: \Cref{sec:analysis} describes the data sample and the analysis methodology, including the correlation function estimators and the treatment of the distortion matrix and metal contamination; \Cref{sec:results} presents the BAO scale parameter measurements, their robustness, and the redshift evolution of the \lya forest and quasar clustering parameters; \Cref{sec:validation} discusses the validation of the analysis pipeline using synthetic datasets; and \Cref{sec:inference} presents the cosmological implications of the distance measurements.
\section{Analysis}\label{sec:analysis}
In this section, we describe the methodology used in our analysis. We rely on the same \lya\ forest data products and quasar catalog as the DESI DR2 \lya\ BAO analysis~\citep{DESI.DR2.BAO.lya}, hereafter \DESIDRIILya. These include the transmitted flux contrast field, $\delta_F$, extracted from over 820,000 quasar spectra, as well as the positions of more than 1.2 million quasars.

All spectral pre-processing steps (continuum fitting, estimation of the flux transmission field, masking of Damped \lya\ Absorbers (DLAs) and Broad Absorption Lines (BALs), and the definition of tracer weights) are identical to those adopted in \DESIDRIILya. A detailed description can be found in previous DESI \lya\ analyses~\citep[e.g.,][]{2023JCAP...11..045G, 2023MNRAS.tmp.3626R, DESI2024.IV.KP6, DESI.DR2.BAO.lya}.

We follow the DR2 analysis for correlation-function estimation and modeling, introducing three modifications to enable BAO measurements at multiple effective redshifts: (i) correlation functions are computed in redshift bins defined at the pair level, (ii) the distortion matrix estimation is adapted to this binning, and (iii) the metal matrices used in the modeling are updated accordingly. These modifications are described below. All other aspects of the pipeline remain unchanged.

\subsection{Correlation Function Estimation}
\label{sec:correlation}
From the DESI DR2 quasar catalog and the catalog of \lya\ transmitted flux contrast $\delta_F$ (see \Cref{eqn:delta_definition}), we compute two correlation functions: the auto-correlation of the \lya\ forest in region A ($1040$–$1205$ \AA\ in the rest frame), hereafter \lyaxlya, and the cross-correlation between \lya\ absorption in region A and quasar positions, hereafter \lyaxqso. We exclude correlation functions derived from region B ($920$–$1020$ \AA) to avoid the substantial increase in covariance matrix size required to account for cross-redshift correlations (see Appendix~\ref{appendix:crossz_cov}). For the full sample, including region B improves the BAO precision by $\sim 10\%$, and while the impact in redshift bins is not yet quantified, it is expected to be comparable. Its inclusion is therefore deferred to future work.

The \lyaxlya\ correlation function, evaluated in bins of transverse and line-of-sight comoving separations $M\equiv(\rp,\rt)$ and in a redshift bin $Z$, is defined as
\begin{equation}
    \xi^{\alpha \times \alpha}_{M,Z}
    = W_{M,Z}^{-1}
      \sum_{\substack{(i,j)\in M \\ z_{\rm pair}\in Z}}
      w^\alpha_i\, w^\alpha_j\, \tilde{\delta}_i\,\tilde{\delta}_j, 
      \label{eq:xiauto}
\end{equation} 
with
\begin{equation}
    W_{M,Z} \equiv 
    \sum_{\substack{(i,j)\in M \\ z_{\rm pair}\in Z}}
    w^\alpha_i\, w^\alpha_j,
    \label{eq:weights}
\end{equation}
where $w^\alpha$ are the \lya\ pixel weights defined as in DESI DR1 and DR2 analyses (Eq.~3.2 of \cite{DESI2024.IV.KP6}). Indices $i$ and $j$ denote individual \lya\ forest pixels, restricted to different quasar spectra so that correlated residuals from the continuum fitting of a single forest do not contribute, and $\tilde{\delta}$ is the projected flux transmission field (see \Cref{subsec:dmat}). The \lyaxqso\ correlation function is defined analogously, replacing one \lya\ pixel weight, $w^\alpha_j$, in \Cref{eq:xiauto,eq:weights} by the quasar tracer weight, $w^Q_j$ (Eq.~3.7 of \cite{DESI2024.IV.KP6}). Unlike the \lyaxlya\ auto-correlation, the estimator contains a single projected flux transmission field, $\tilde{\delta}_i$, associated with the \lya\ forest.

Each pixel-pixel or pixel-quasar pair is assigned to a redshift bin according to its mean pair redshift,
\begin{equation}
    z_{\rm pair} \equiv \tfrac{1}{2}(z_i + z_j).
\end{equation}
This choice differs from the approach adopted in eBOSS~\citep{deSainteAgathe:DR14, Blomqvist:DR14, dmdB:DR16}, where correlations were binned at the forest level rather than at the pixel level. In that framework, forest-forest pairs were assigned a redshift bin using the average of the maximum usable redshifts of the two forests, while quasar-forest pairs used the average of the forest maximum redshift and the quasar redshift. This strategy ensured that individual forests were not split across redshift bins, thereby limiting the propagation of continuum-fitting distortions between bins. In contrast, our pixel-based binning allows a single forest to contribute to multiple redshift bins. This implies that continuum-fitting distortions can, in principle, induce correlations between redshift bins and requires a more detailed treatment of the distortion matrix. We discuss this in more detail in \Cref{subsec:dmat}.

We divide the sample into three redshift bins: $z_{\rm pair} \leq 2.25$, $2.25 < z_{\rm pair} \leq 2.6$, and $z_{\rm pair} > 2.6$. The number of bins is chosen to balance redshift resolution and statistical precision, ensuring that each bin retains sufficient information for a robust BAO measurement while still allowing for the study of the redshift evolution of our model parameters. The bin boundaries are chosen to provide comparable statistical power across bins and to ensure a redshift extent of at least $300\,\hMpc$ adopted in \DESIDRIILya\ for distortion matrix estimation.\footnote{As shown in \citep{Busca:dmat}, this extent is sufficient for the auto-correlation but not for the cross-correlation, which would ideally require $\sim400\,\hMpc$; we retain $300\,\hMpc$ for consistency with \DESIDRIILya\ and leave a larger extent to future work.}

\begin{figure}[!tbp]
    \centering
    \includegraphics[width=0.99\textwidth]{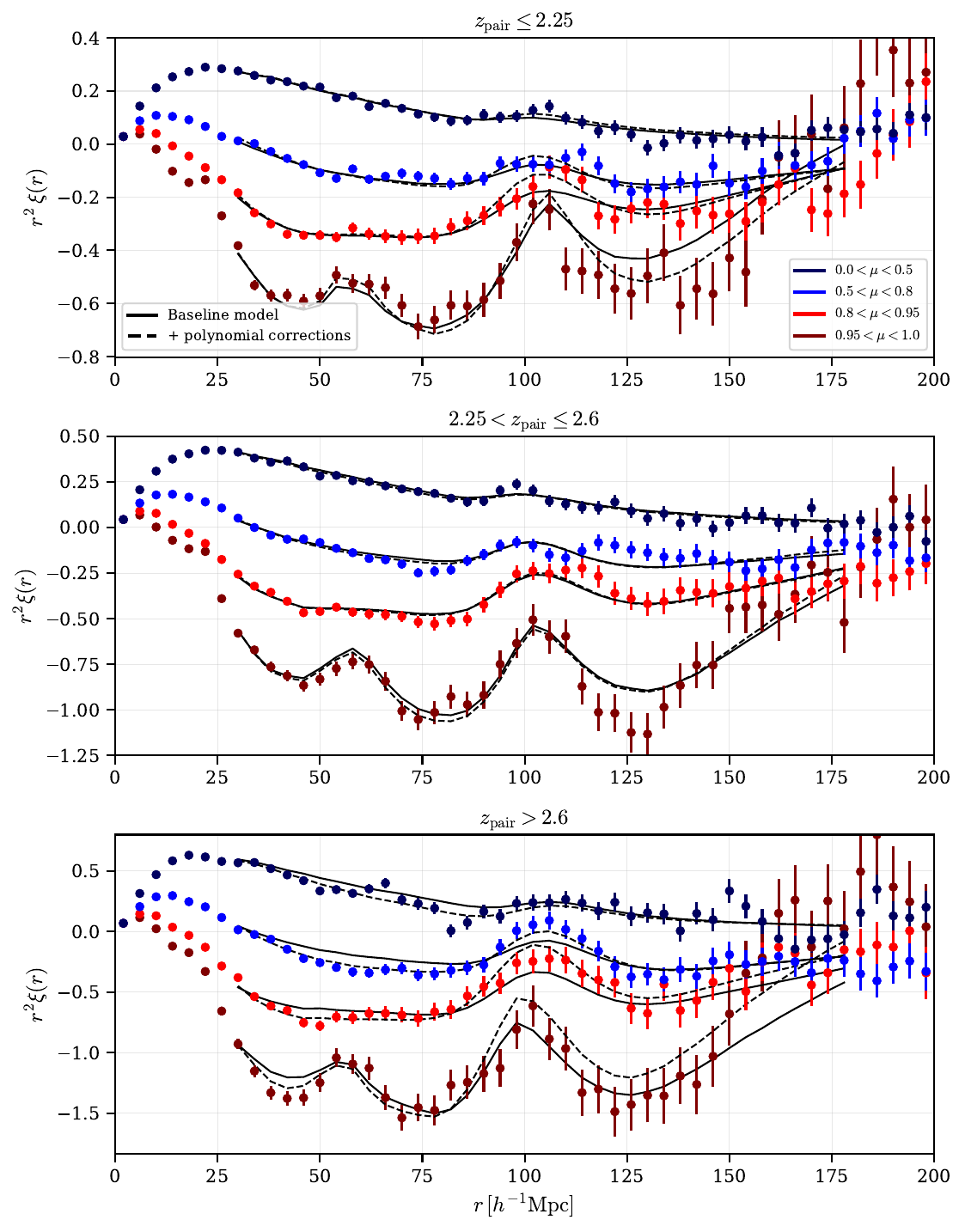}
    \caption{Measured \lyaxlya\ auto-correlation function in three redshift bins defined by $z_{\rm pair} \leq 2.25$ (top panel), $2.25 < z_{\rm pair} \leq 2.6$ (middle panel), and $z_{\rm pair} > 2.6$ (bottom panel). Colors indicate the orientation of the separation vector interval with respect to the line of sight, with brown corresponding to pairs closer to the line of sight and blue to more transverse separations. Solid black lines show the best-fit baseline model (see \Cref{subsec:best-fit}), while dashed lines correspond to the best-fit model including additive broadband polynomial corrections (see \Cref{subsec:goodnessoffit}).}
    \label{fig:corrs_auto}
\end{figure}

\begin{figure}[!tbp]
    \centering
    \includegraphics[width=\textwidth]{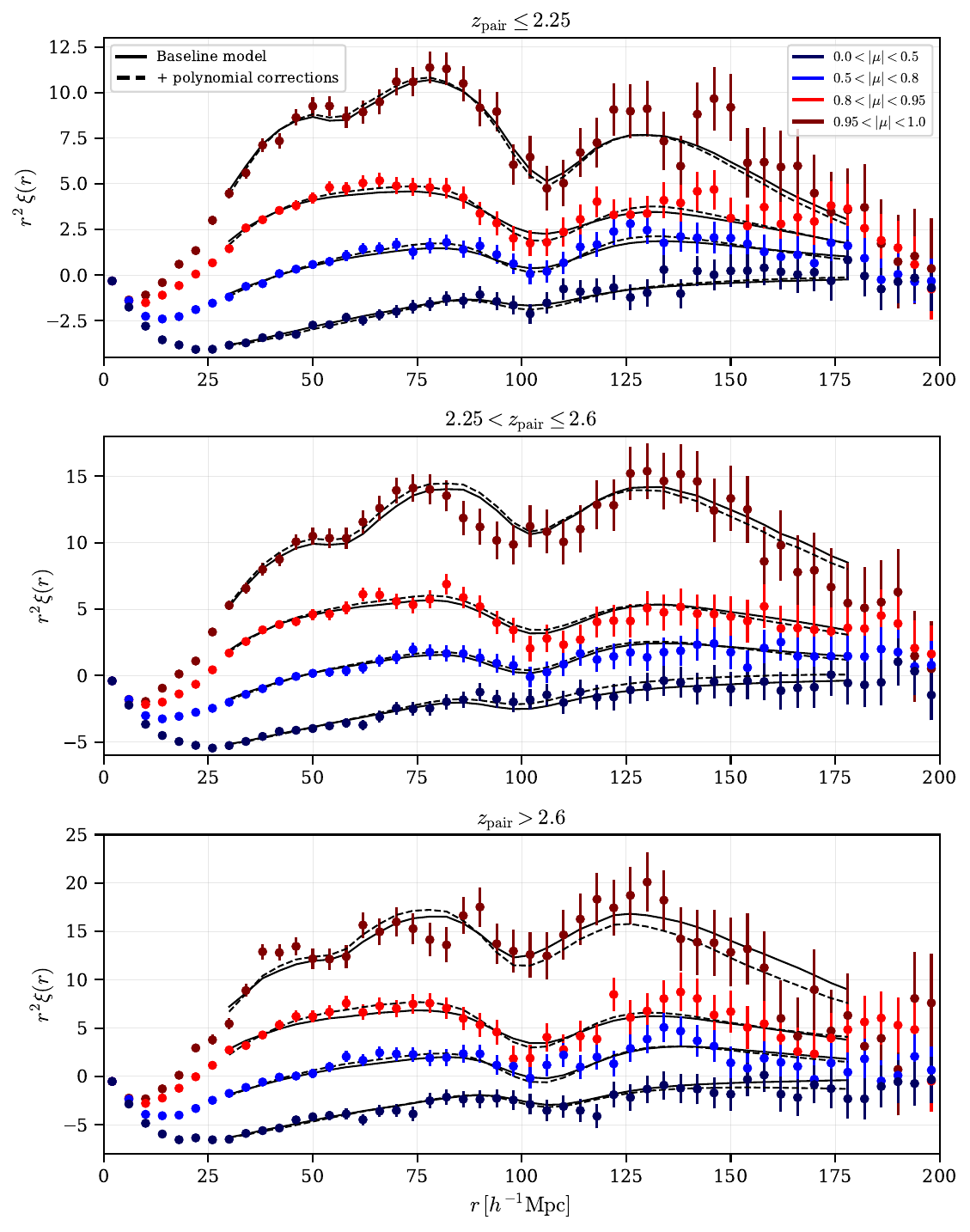}
    \caption{As in \Cref{fig:corrs_auto}, but for the \lyaxqso\ correlation function.}
    \label{fig:corrs_cross}
\end{figure}

The correlation functions are computed using the \texttt{picca} software~\citep{picca}\footnote{\url{https://github.com/igmhub/picca}}. Following standard practice, results are presented as a function of $r = \sqrt{r_\parallel^2 + r_\perp^2}$ in wedges defined by $\mu = r_\parallel / r$. The measurements are shown in \Cref{fig:corrs_auto,fig:corrs_cross}.

\subsection{Distortion Matrix}\label{subsec:dmat}

The \lya\ transmitted flux field is defined as
\begin{equation}
    \delta_{q}\left(\lambda\right) = \frac{f_{q}\left(\lambda\right)}{\overline{F}\left(\lambda\right)C_{q}\left(\lambda\right)} - 1~,
    \label{eqn:delta_definition}
\end{equation}
where $f_q(\lambda)$ is the flux from a quasar $q$ as a function of wavelength $\lambda$, $\overline{F}\left(\lambda\right)$ is the mean transmission, and $C_{q}\left(\lambda\right)$ is the unabsorbed quasar continuum. Ideally, the quantity $\delta_{q}$ would be used to compute correlation functions. However, in this work, as in \DESIDRIILya, the continuum estimation method, based on a fit to the same data, mixes absorption from different wavelengths, distorting the measured correlation function with respect to that of the original field~\citep{Busca:dmat}.

This effect is addressed by explicitly applying a linear projection operator to the measured transmitted flux field, yielding a projected field whose correlation properties match those of the true projected field provided that continuum-template imperfections are uncorrelated between different forests. The advantage of this approach is that the resulting distortion of the correlation function can be computed explicitly and incorporated into the theoretical model used in the data analysis.

The projected field $\tilde \delta_q$ is defined by the linear combination 
$\tilde \delta_{q,i} = \sum_j \eta_{ij} ~ \delta_{q,j}$,
where $i$ and $j$ are wavelength indices from the same quasar spectrum $q$, and the coefficients $\eta_{ij}$ are given in Eq. 3.3 of \cite{DESI2024.IV.KP6}. From Eq.~\ref{eq:xiauto}, we obtain for the separation bin $M$ and redshift bin $Z$, 
\begin{eqnarray}
\tilde \xi_{M,Z} &=& W^{-1}_{M,Z} \sum_{\substack{(i,j)\in M \\ z_{\rm pair}\in Z}} w^\alpha_i w^\alpha_j \sum_{k,p} \eta_{i,k} \eta_{j,p} \delta_k \delta_p\,.
\end{eqnarray}

Inserting another set of separation bins $N$, we obtain on average,
\begin{eqnarray}
\tilde \xi_{M,Z}            
          &=& \sum_N \left[ W^{-1}_{M,Z} \sum_{\substack{(i,j)\in M \\ z_{\rm pair}\in Z}} w^\alpha_i w^\alpha_j \sum_{(k,p) \in N}  \eta_{i,k} \eta_{j,p} f(z_k,z_p,z_{\rm ref}) \right] \xi_{N,z_{\rm ref}}\,,
\end{eqnarray}
where $\xi_{N,z_{\rm ref}}$ is the undistorted correlation function in the separation bin $N$ evaluated at a reference redshift $z_{\rm ref}$, and the function $f(z_k,z_p,z_{\rm ref})$ accounts for the evolution of the growth of structure between $z_{\rm ref}$ and the redshifts of each measurement (see \DESIDRIILya). 

The quantity in square brackets is the $(M,N)$ element of the distortion matrix relating the undistorted correlation function evaluated at $z_{\rm ref}$ to the distorted correlation measured in redshift bin $Z$. This expression is identical to that used in \DESIDRIILya, except that the sum over $(i,j)$ pairs is restricted to those assigned to the selected pair redshift bin. Although the measured correlation function is computed using only pixel pairs within a given redshift bin, the projected flux field of each pixel is a linear combination of all pixels in the corresponding forest. Consequently, a pair assigned to one redshift bin can receive contributions from absorption at wavelengths associated with another bin, allowing the distortion matrix to couple different redshift bins. This coupling is absent in the forest-level binning adopted in eBOSS, where each forest belongs to a single redshift bin. We quantify the resulting cross-bin correlations in Appendix~\ref{appendix:crossz_cov}. An analogous expression applies to the \lyaxqso\ cross-correlation, with the \lya\ weight $w^\alpha_j$ replaced by the quasar weight $w^Q_j$, while the projection is applied only to the \lya\ field.

\subsection{Contamination by Metals}
Absorption by other atomic transitions in the intergalactic medium, hereafter referred to as metals, introduces spurious correlations that can contaminate the \lya\ auto- and cross-correlation measurements if unaccounted for. Because the analysis assumes all absorption arises from \lya, metals at different rest-frame wavelengths are misinterpreted as \lya\ at incorrect redshifts, producing displaced, non-physical correlations along the line of sight.

Following \citep{Blomqvist:2018}, these effects are accounted for using metal matrices, $M_{AB}$, where $A$ and $B$ denote comoving separation bins; \DESIDRIILya\ improved this treatment by modeling the variation of the metal correlations with $r_\perp$. Each matrix encodes the mapping between the true separation (set by the metal transition) and the measured separation under the \lya\ assumption. We generalize this approach to pair-based redshift binning by including only pairs whose mean assumed redshift falls within the bin of interest.

\section{Results}\label{sec:results}

In this section, we present the BAO measurements obtained from the \lya\ auto- and cross-correlation functions in the three redshift bins defined in \Cref{sec:analysis}. We first report the BAO parameters and assess their fit quality and robustness, and then present the redshift evolution of the effective \lya\ forest bias and RSD parameters, together with the quasar bias, as obtained from the joint multi-bin fit. The remaining nuisance parameters and their redshift dependence are discussed in Appendix~\ref{appendix:nuisance}.

\subsection{Constraints on BAO scale parameters}\label{subsec:best-fit}

The measured correlation functions are fitted using the baseline BAO model of the DESI DR2 \lya\ BAO analysis, implemented with the \texttt{Vega} package\footnote{\url{https://github.com/andreicuceu/vega}}. The model includes two BAO scale parameters, $\alpha_\parallel$ and $\alpha_\perp$, which quantify shifts in $D_H/r_d$ and $D_M/r_d$ relative to their fiducial values. In addition, 15 nuisance parameters are considered, including the \lya\ forest bias $b_\alpha$, the redshift-space distortion parameter $\beta_\alpha$, and the quasar bias $b_Q$, along with components accounting for metal contamination, correlated noise induced by the data processing pipeline~\citep{KP6s5-Guy}, high-column density systems~\citep[HCDs;][]{Tan:2025}, quasar redshift errors and peculiar velocities~\citep{KP6s4-Bault}, and the transverse proximity effect~\citep{Hada:2026}. Section~IV of \DESIDRIILya\ and Section~4 of \cite{DESI2024.IV.KP6} present a more detailed description of the modeling framework.

We adopt a Gaussian likelihood. Best-fit parameters and goodness-of-fit statistics are obtained using the \texttt{iminuit} minimizer~\citep{James:1975dr,iminuit}, while posterior distributions are sampled with the \texttt{Polychord} nested sampler~\citep{Handley:2015fda,Handley:2015Polychord}, both interfaced through \texttt{Vega}. Priors follow those of the DESI DR2 \lya\ BAO analysis (Table~II of \DESIDRIILya), except for the quasar bias, which is assigned a uniform prior to allow redshift evolution. The resulting BAO scale parameters from the combined fit of the auto- and cross-correlation are listed in \Cref{tab:BAOparams}. The scale parameters are defined as
\begin{equation}
    \alpha_\parallel = \frac{D_H(\zeff)/r_d}{\left[D_H(\zeff)/r_d\right]^{\rm fid}}, \qquad
    \alpha_\perp = \frac{D_M(\zeff)/r_d}{\left[D_M(\zeff)/r_d\right]^{\rm fid}},
    \label{eq:alpha_def}
\end{equation}
where the superscript ``fid'' denotes the fiducial cosmology. Throughout this work, we adopt the flat $\Lambda$CDM Planck 2018 cosmology~\citep{Planck:2018vyg_cosmological-parameters} used in \DESIDRIILya\ (see their Table~I). In addition to $\alpha_\parallel$ and $\alpha_\perp$, we report the derived isotropic and anisotropic (Alcock-Paczyński) dilation parameters,
\begin{equation}
    \alpha_{\rm ISO} = \alpha_\parallel^{0.55}\,\alpha_\perp^{0.45}, \qquad
    \alpha_{\rm AP} = \frac{\alpha_\perp}{\alpha_\parallel},
    \label{eq:alphaiso_ap}
\end{equation}
following \DESIDRIILya.

Our baseline model includes a term describing non-linear broadening of the BAO peak and a small-scale correction for the auto-correlation following~\citep{Arinyo2015}. The corresponding parameters are fixed at the effective redshift of each bin: the broadening is set using Lagrangian Perturbation Theory predictions~\citep{2007EisensteinBroadening,2013KirkbyFitting}, and the small-scale correction is obtained via interpolation from the ACCEL2 simulation suite~\citep{Chabanier2024}. The impact of these assumptions is assessed in \Cref{subsec:robustness}.

\begin{table}[t]
\centering
\begin{tabular}{llll}

Parameter  &  $z_{\rm pair} \leq 2.25$ & $2.25 < z_{\rm pair} \leq 2.6$ & $z_{\rm pair} > 2.6$  \\
\hline
$\alpha_{\parallel}$   & $0.9997\pm0.022$           & $0.990\pm0.020$            & $1.016\pm0.024$ \\
$\alpha_{\perp}$ & $0.995\pm0.022$            & $1.021\pm0.025$            & $0.946^{+0.022}_{-0.025}$ \\ 
$\rho_{\alpha_{\parallel},\alpha_{\perp}}$ & $-0.469$ & $-0.467$ & $-0.490$ \\
\hdashline
$\alpha_{\rm ISO}$         & $0.998\pm0.011$            & $1.004\pm0.011$            & $0.984\pm0.012$ \\
$\alpha_{\rm AP}$          & $0.996\pm0.037$            & $1.032\pm0.040$            & $0.931^{+0.036}_{-0.041}$ \\
\hline
$z_{\rm eff}$
& $2.13$
& $2.40$
& $2.81$ \\
$\chi_{\rm min}^2$ 
& $4849.24$ 
& $4612.68$ 
& $4750.15$ \\
DoF
& $4653-17$ 
& $4653-17$ 
& $4653-17$ \\
PTE 
& $0.01$ 
& $0.59$ 
& $0.12$ \\
\hline
\end{tabular}
\caption{
Marginalized posterior constraints on the BAO parameters $\alpha_\parallel$ and $\alpha_\perp$, and correlation coefficient $\rho$ from combined fits to the \lya\ auto- and cross-correlation functions in three redshift bins defined by the mean pair redshift $z_{\rm pair}$ interval. Quoted uncertainties correspond to $1\sigma$ credible intervals. The dashed rule separates the derived isotropic dilation parameter $\alpha_{\rm ISO}$ and the Alcock-Paczyński parameter $\alpha_{\rm AP}$, defined in \Cref{eq:alphaiso_ap} following \DESIDRIILya. The table also lists the effective redshift $z_{\rm eff}$, the minimum $\chi^2$, the number of degrees of freedom (DoF), and the probability to exceed (PTE) for each bin.
}

\label{tab:BAOparams}
\end{table}

\subsection{Goodness of fit}\label{subsec:goodnessoffit}
The $\chi^2$ values and associated probabilities to exceed (PTE) reported in \Cref{tab:BAOparams} quantify the agreement between model and data over the full fitting range, $30 < r < 180\ h^{-1}\mathrm{Mpc}$. The intermediate and highest redshift bins yield acceptable PTE values, while the lowest redshift bin shows a comparatively low value ($\mathrm{PTE}=0.01$). Since $\chi^2$ reflects the agreement across the entire fitting range, such a value can arise from localized discrepancies unrelated to the BAO feature itself. We investigate its origin below.

\begin{figure}[!tbp]
    \centering
    \includegraphics[width=0.95\linewidth]{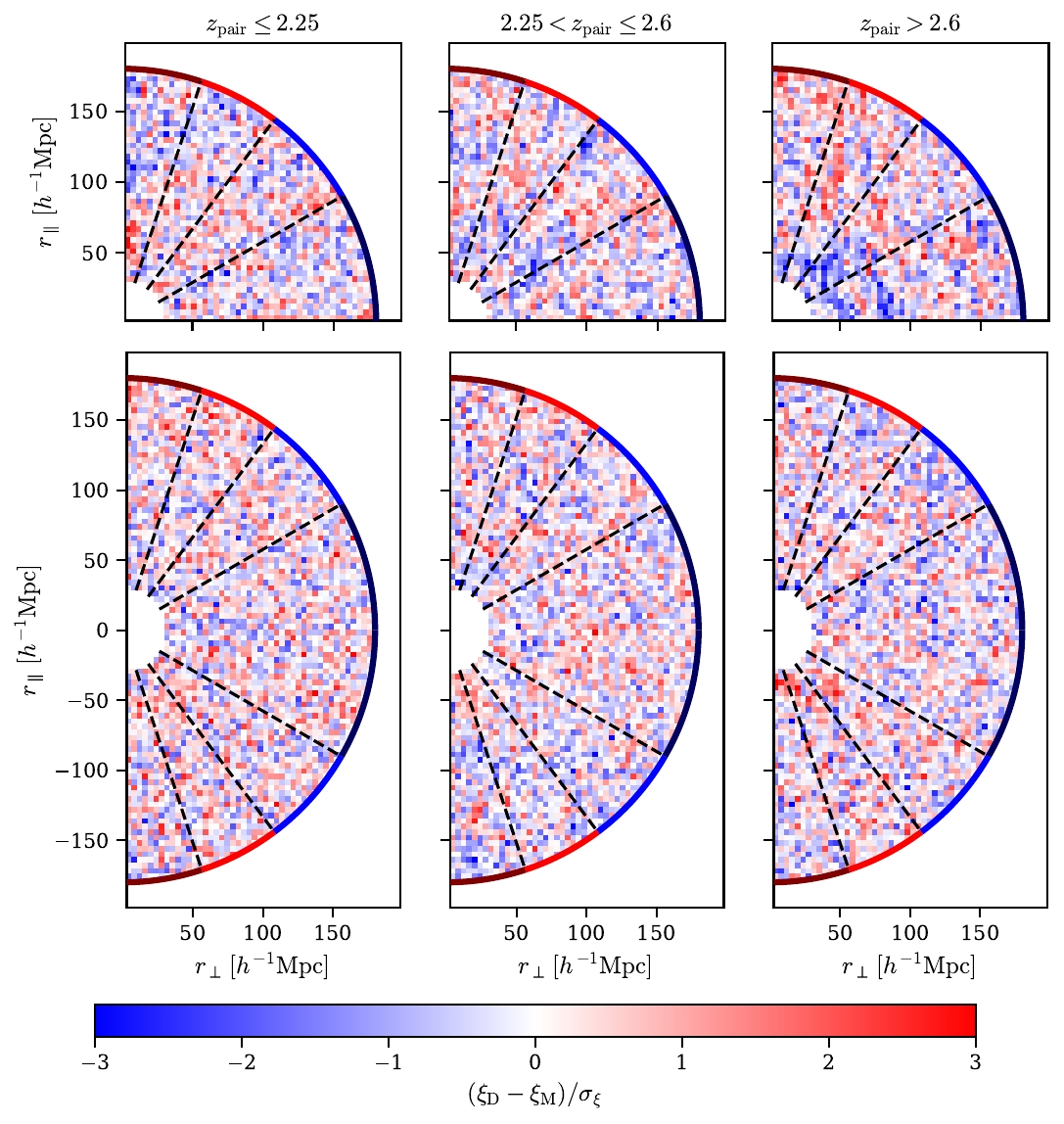}
    \caption{Normalized residuals, $(\xi_{\rm D} - \xi_{\rm M}) / \sigma_\xi$, for the auto-correlation (top row) and cross-correlation (bottom row) as a function of $r_\parallel$ and $r_\perp$ in the three redshift bins (columns). The color scale is saturated at $\pm3\sigma$. Colored arcs at $r = 180\,\hMpc$, separated by black dashed dividers, mark the boundaries of the $\mu$ wedges ($|\mu| = 0.5$, $0.8$, $0.95$) used in \Cref{fig:corrs_auto,fig:corrs_cross}, with colors matching those figures.}
    \label{fig:residuals}
\end{figure}

From \Cref{fig:corrs_auto,fig:corrs_cross}, small mismatches between the baseline model and the data are visible. However, the wedge representation can obscure localized features, as it averages over angular bins. To examine this in more detail, \Cref{fig:residuals} shows the normalized residuals $(\xi_{\rm D}-\xi_{\rm M})/\sigma_\xi$ in the $(r_\parallel,r_\perp)$ plane.

As seen in \Cref{fig:residuals}, most residuals are consistent with noise. The largest deviations occur in the lowest redshift bin, where the auto-correlation shows a coherent structure at small transverse separations ($r_\perp \lesssim 20\ h^{-1}\mathrm{Mpc}$), with a positive excess near $r_\parallel \sim 50\ h^{-1}\mathrm{Mpc}$ and a negative feature around $r_\parallel \sim 100\ h^{-1}\mathrm{Mpc}$. Similar, though weaker, patterns are present in other bins, and the cross-correlation exhibits a feature in the highest redshift bin near $r_\parallel \sim -40\ h^{-1}\mathrm{Mpc}$. These structures are consistent with spurious correlations induced by quasar redshift errors propagating through the continuum-fitting procedure~\citep{Youles:RedshiftErrors, Gordon:RedshiftErrors}.

Additional negative features appear in the auto-correlation of the highest redshift bin at $r_\perp \sim 50$ and $75\ h^{-1}\mathrm{Mpc}$. We do not identify a clear origin for these structures, and they may reflect statistical fluctuations.

Motivated by these observations, we perform a series of tests:
\begin{itemize}[noitemsep]
    \item First, we repeat the analysis including an additive broadband polynomial correction to the model (dashed lines in \Cref{fig:corrs_auto,fig:corrs_cross}). The correction is described by Legendre polynomials $L_j(\mu)$ with $j=0,2,4,6$ multiplied by powers of $r^{-i}$ with $i=0,1,2$, introducing 12 additional free parameters per correlation function (24 in total). Despite this increased flexibility, the PTE of the lowest redshift bin remains low ($\mathrm{PTE}=0.02$), while the intermediate and highest redshift bins yield $\mathrm{PTE}=0.59$ and $\mathrm{PTE}=0.21$.

    \item Second, we restrict the fitting range to $60 < r < 160\ h^{-1}\mathrm{Mpc}$ (compared to the baseline $30 < r < 180\ h^{-1}\mathrm{Mpc}$). This raises the PTE of the lowest-redshift bin from $0.01$ to $0.11$, while the intermediate- and highest-redshift bins remain acceptable, with $\mathrm{PTE}=0.46$ and $0.17$, respectively. The improvement in the lowest-redshift bin suggests that scales below $60\ h^{-1}\mathrm{Mpc}$ contribute to the low PTE observed in the baseline fit.

    \item Third, we restrict the $\mu$ range to $\mu < 0.95$ for the auto-correlation and $\mu > -0.95$ for the cross-correlation, thereby reducing the contribution from near line-of-sight separations. This yields PTE values of $\mathrm{PTE}=0.19$, $\mathrm{PTE}=0.79$, and $\mathrm{PTE}=0.28$ for the three redshift bins, showing that these configurations contribute significantly to the low PTE.

    \item Finally, we repeat the analysis discarding close pairs. These are defined as pixel–pixel pairs where the forests are angularly close and the velocity separation between each pixel and the background quasar of the other forest falls below a threshold, and pixel–quasar pairs where both angular and velocity separations between the forest quasar and the quasar tracer fall below the same threshold. The thresholds are chosen to correspond to a comoving separation of $15\, h^{-1}\mathrm{Mpc}$ at the effective redshift of each bin. Such pairs are expected to be most affected by spurious correlations induced by quasar redshift errors~\citep{Youles:RedshiftErrors, Gordon:RedshiftErrors, Y3.lya-s1.Casas.2025}. This yields PTE values of $0.03$, $0.56$, and $0.15$, indicating only a modest improvement.
\end{itemize}

These tests indicate that the low PTE is driven primarily by discrepancies on small scales and near the line of sight, rather than by a global failure of the model. Several effects could contribute to these discrepancies. Along the line of sight, possible contributions include the distortion induced by continuum fitting, whose impact is strongest in this direction~\citep{Busca:dmat}, as well as residual quasar-redshift errors or an imperfect treatment of HCD systems. On small scales, unmodeled contaminants such as metal transitions not included in the baseline model, or residual instrumental systematics, may also play a role. We note, however, that the tests restricting the fitting range in $r$ and $\mu$ (the second and third tests above) also reduce the number of data points entering the $\chi^2$ evaluation, which by itself can increase the PTE.

As shown in \Cref{sec:validation}, a further contribution comes from covariance-matrix smoothing, which shifts the $\chi^2$ distribution toward higher values; accounting for this effect raises the PTE values to $0.08$, $0.90$, and $0.26$, with negligible impact on the BAO uncertainties ($\lesssim 1.15\%$).

We therefore attribute the low PTE in the lowest-redshift bin to a combination of localized residual structure and covariance-related effects, which become more noticeable due to the smaller \lya\ clustering amplitude in this bin, thereby increasing their relative contribution. This reflects the higher sensitivity of the lowest-redshift bin rather than a failure of the BAO model. Importantly, as shown in \Cref{subsec:robustness}, the recovered BAO scale remains stable under all the tests considered here.

\subsection{Robustness of the BAO measurements}\label{subsec:robustness}
\begin{figure}[!tbp]
    \centering
    \includegraphics[width=\linewidth]{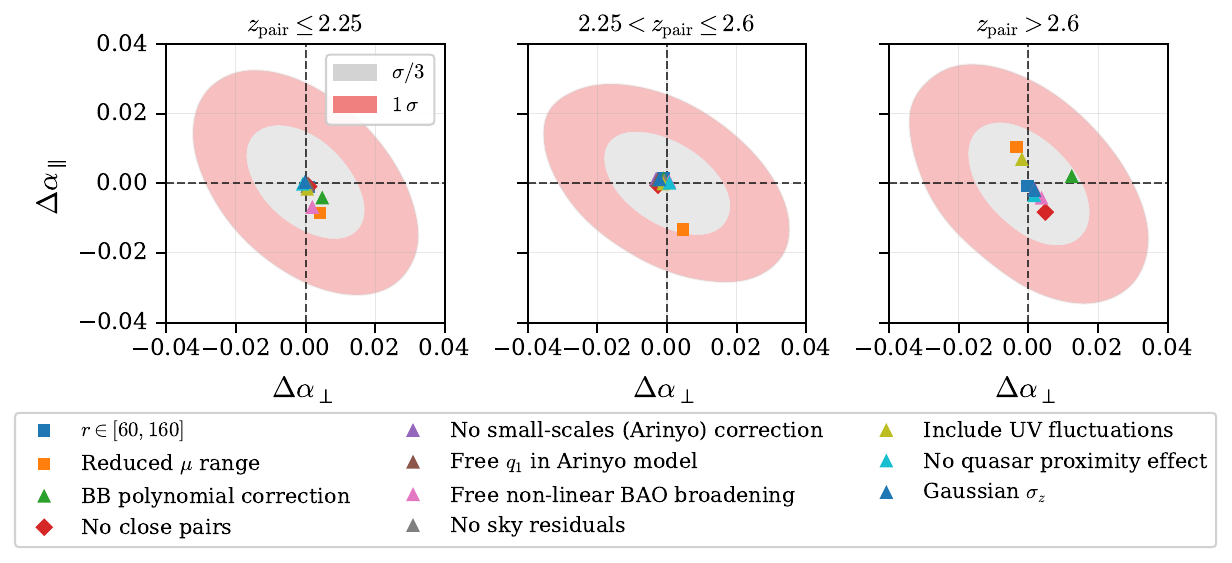}
    \caption{Shifts in the BAO scale parameters $\Delta \alpha_\perp$ and $\Delta\alpha_\parallel$ relative to the baseline fit, shown for each redshift bin. Filled contours indicate the $\sigma/3$ (grey) and $1\sigma$ (red) confidence regions of the baseline. Markers show the best-fit shifts for each systematic variation, color-coded by variant and grouped by type: squares for fitting-range variations, triangles for modeling variations, and a diamond for the close-pair removal test.}
    \label{fig:robustness}
\end{figure}

We assess the stability of the BAO measurements under variations in the analysis strategy, fitting configuration, modeling assumptions, and prior choices. This complements the goodness-of-fit analysis and quantifies the impact of the identified residual features on the recovered BAO scale.

Following \DESIDRIILya, we adopt a threshold of $\sigma/3$ on shifts in $\alpha_\parallel$ and $\alpha_\perp$ relative to the baseline. The results are summarized in \Cref{fig:robustness}, including the fitting range, broadband correction, and close-pair removal tests discussed in \Cref{subsec:goodnessoffit}, along with the additional variations described below.

\paragraph{Model assumptions.}
We consider variations including: freeing the BAO broadening parameters; removing the small-scale correction; freeing $q_1$, the parameter controlling the amplitude of the non-linear small-scale correction~\citep{Arinyo2015}, instead of fixing it; ignoring correlated sky residuals; excluding the transverse proximity effect; modeling quasar redshift errors with a Gaussian rather than a Lorentzian; and including UV background fluctuations.

\paragraph{Prior choices.}
We also test the impact of informative priors by removing, one at a time, those on the HCD RSD parameter and characteristic scale ($\beta_{\rm HCD}$ and $L_{\rm HCD}$), the systematic line-of-sight shift from quasar redshift errors ($\Delta r_\parallel$), and the C\,\textsc{IV} bias ($b_{\rm CIV}$). These variations produce shifts below $0.01\sigma$ and are not shown in \Cref{fig:robustness} for clarity.

All tested variations yield shifts within the $\sigma/3$ threshold in all redshift bins, indicating that the BAO measurements are insensitive to reasonable changes in modeling and analysis choices.

\begin{figure}[!tbp]
    \centering
    \includegraphics[width=\linewidth]{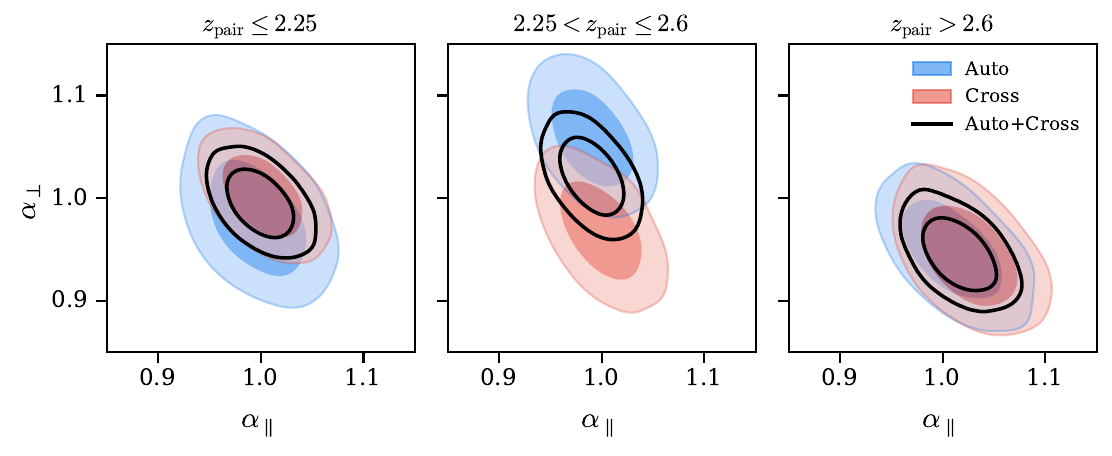}
    \caption{BAO scale parameters $\alpha_\parallel$ and $\alpha_\perp$ from separate fits to the \lya\ auto-correlation (blue) and cross-correlation (red), and the auto+cross baseline analysis (black, unfilled), in the three redshift bins. Contours show the $1\sigma$ and $2\sigma$ credible regions.}
    \label{fig:auto_cross}
\end{figure}

As an additional consistency check, we perform independent fits to the auto- and cross-correlation functions; the resulting BAO constraints are shown in \Cref{fig:auto_cross}. While the lowest- and highest-redshift bins show good agreement between the two measurements ($<0.5\sigma$ in both BAO parameters), the middle redshift bin shows a statistically marginal offset of $2.0\sigma$ in $\alpha_\perp$; the corresponding $\alpha_\parallel$ values agree to within $0.3\sigma$. Consistently, a smaller offset in the same direction ($0.7\sigma$) was observed in the single-redshift measurement (Figure~10 and Table~II of \DESIDRIILya). Given its marginal significance, we interpret this difference as consistent with a statistical fluctuation.

\subsection{Redshift evolution of \lya\ forest and quasar bias}\label{subsec:nuisance_evol}

\begin{figure}[!tbp]
    \centering
    \includegraphics[width=\linewidth]{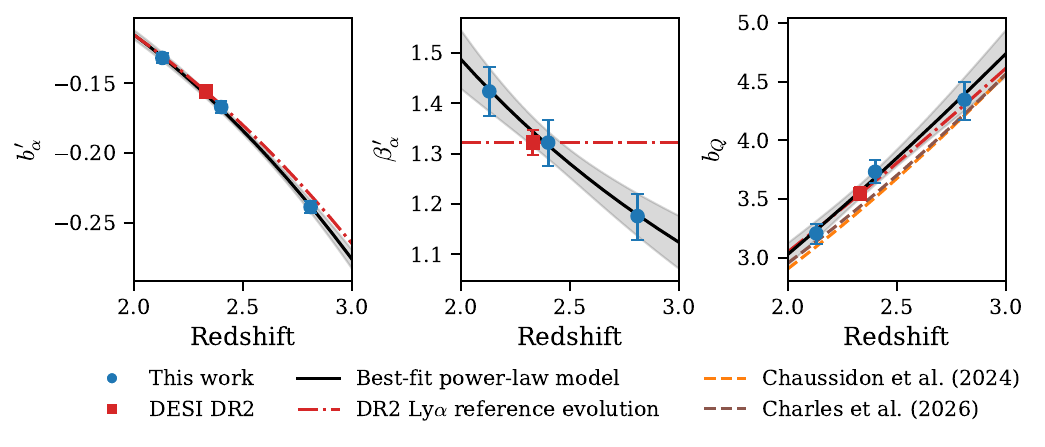}
    \caption{Redshift evolution of the effective \lya\ bias $b_\alpha'$ (left), the RSD parameter $\beta_\alpha'$ (center), and the quasar bias $b_Q$ (right). Blue circles show our measurements in the three redshift bins; red squares show the corresponding \DESIDRIILya\ value at $z_{\rm eff} = 2.33$. Error bars denote $1\sigma$ uncertainties. Solid black curves show best-fit power-law models with $1\sigma$ bands. Red dash-dotted curves indicate assumed scalings in the baseline model. In the right panel, independent quasar bias measurements from DESI DR1 and DR2 clustering analyses~\citep{ChaussidonY1fnl, Charles:2026} are shown for comparison.}
    \label{fig:nuisance_evol}
\end{figure}

As stated in \Cref{sec:analysis}, the purpose of dividing the sample into three redshift bins in this work is, in addition to obtaining tight BAO constraints, to enable a direct measurement of the redshift evolution of the parameters in our model within a single self-consistent analysis. \Cref{fig:nuisance_evol} shows the evolution of the effective \lya\ forest bias $b_\alpha'$, the redshift-space distortion (RSD) parameter $\beta_\alpha'$, and the quasar bias $b_Q$ across the three redshift bins, compared to the single-bin results from \DESIDRIILya. The full set of nuisance-parameter marginalized constraints is reported in Appendix~\ref{appendix:nuisance}.

The \lya\ bias $b_\alpha$ and RSD parameter $\beta_\alpha$ are strongly degenerate with the high-column-density (HCD) bias $b_{\rm HCD}$. We therefore quote the effective quantities~\citep{Font-Ribera:2012}, defined as
\begin{align}
b_\alpha' &= b_\alpha + b_{\rm HCD}\,F_{\rm HCD}(k_\parallel), \label{eq:beff}\\
b_\alpha'\beta_\alpha' &= b_\alpha\beta_\alpha + b_{\rm HCD}\beta_{\rm HCD}\,F_{\rm HCD}(k_\parallel), \label{eq:betaeff}
\end{align}
where $F_{\rm HCD}(k_\parallel) = \exp(-L_{\rm HCD}\,k_\parallel)$ is a function related to the number and column-density distribution of HCDs~\citep{Rogers:2018}. We adopt the large-scale limit $k_\parallel\to0$, in which $F_{\rm HCD}\to1$ and $L_{\rm HCD}$ drops out of \Cref{eq:beff,eq:betaeff}, so that $b_\alpha'$ and $\beta_\alpha'$ combine the \lya\ and HCD contributions into single effective quantities that can be constrained independently of the \lya--HCD degeneracy. In practice $F_{\rm HCD}$ should be evaluated at an effective $k_\parallel$. However, determining it requires a dedicated study of the HCD model over the fitted scales in each redshift bin, which we leave to future work.

We quantify the evolution of each parameter with the power-law model
\begin{equation}
    X(z) = X_0 \left(\frac{1+z}{1+z_0}\right)^{\gamma_X},
    \label{eq:powerlaw_evol}
\end{equation}
with pivot $z_0 = 2.33$, fitted independently to each parameter. Cross-bin covariance is neglected for all parameters except the quasar bias $b_Q$, as justified in Appendix~\ref{appendix:crossz_cov}.

The effective bias $b_\alpha'$ becomes more negative with increasing redshift, as expected for the \lya\ forest tracing increasingly neutral gas at earlier times. The fit gives $b_{\alpha,0}' = -0.1581 \pm 0.0026$ and $\gamma_\alpha = 3.05 \pm 0.16$, consistent within $1\sigma$ with the \DESIDRIILya\ value $b_\alpha' = -0.1558 \pm 0.0027$ and with the commonly adopted scaling $\gamma_\alpha = 2.9$. The RSD parameter $\beta_\alpha'$ decreases with redshift, with $\beta_{\alpha,0}' = 1.343 \pm 0.029$ and $\gamma_\beta = -0.97 \pm 0.26$. This corresponds to a $\sim3.7\sigma$ preference for non-zero evolution, and a similar redshift dependence is also observed in hydrodynamical simulations~\citep[e.g.,][]{Arinyo2015,Chabanier2024}. However, the impact of DLA masking on this measurement has not yet been fully assessed, and a dedicated study will be required to confirm the robustness of the observed evolution. The quasar bias $b_Q$ increases with redshift, with $b_{Q,0} = 3.562 \pm 0.059$ and $\gamma_Q = 1.56 \pm 0.23$, in agreement with the standard scaling $\gamma_Q = 1.44$~\citep{2019ApJ...878...47D} and with independent DESI quasar clustering measurements~\citep{ChaussidonY1fnl, Charles:2026}, providing a cross-check across different observables.

\section{Validation with Synthetic Spectra}\label{sec:validation}

We validate our analysis pipeline using the synthetic datasets constructed for the DESI DR2 \lya\ BAO analysis~\citep{Y3.lya-s1.Casas.2025}. This provides an independent assessment of the conclusions drawn in \Cref{subsec:goodnessoffit,subsec:robustness}, in particular the robustness of the BAO measurements and the interpretation of the $\chi^2$ and PTE values. We apply our pair-based redshift binning procedure independently to each mock realization. The mock suite consists of 300 realizations of the \texttt{CoLoRe-QL} mocks~\citep{Farr:2020, Y3.lya-s1.Casas.2025} and 100 realizations of the \texttt{Saclay} mocks~\citep{Etourneau:2024}. A detailed description of the synthetic spectra generation pipeline and the mock datasets is presented in~\citep{2024arXiv240100303H, Y3.lya-s1.Casas.2025}.

For each realization, we measure the \lya\ auto- and cross-correlation functions in the three redshift bins defined in \Cref{sec:analysis} and fit for the BAO scale parameters $\alpha_\parallel$ and $\alpha_\perp$ following the same procedure as for the data. In addition, we stack the correlation functions across all realizations of each mock suite to obtain high signal-to-noise measurements, which we use to test for systematic biases in the pipeline.

\begin{figure}[!tbp]
    \centering
    \includegraphics[width=\linewidth]{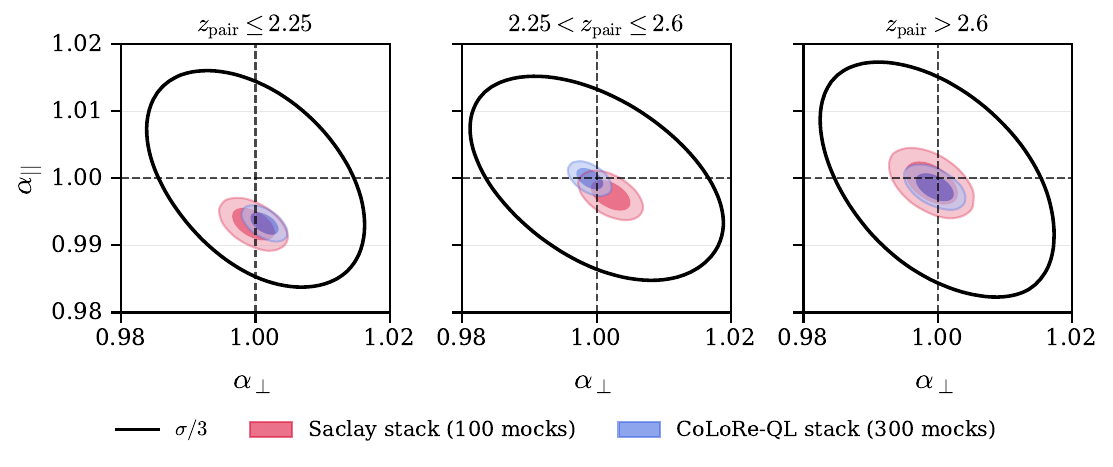}
    \includegraphics[width=\linewidth]{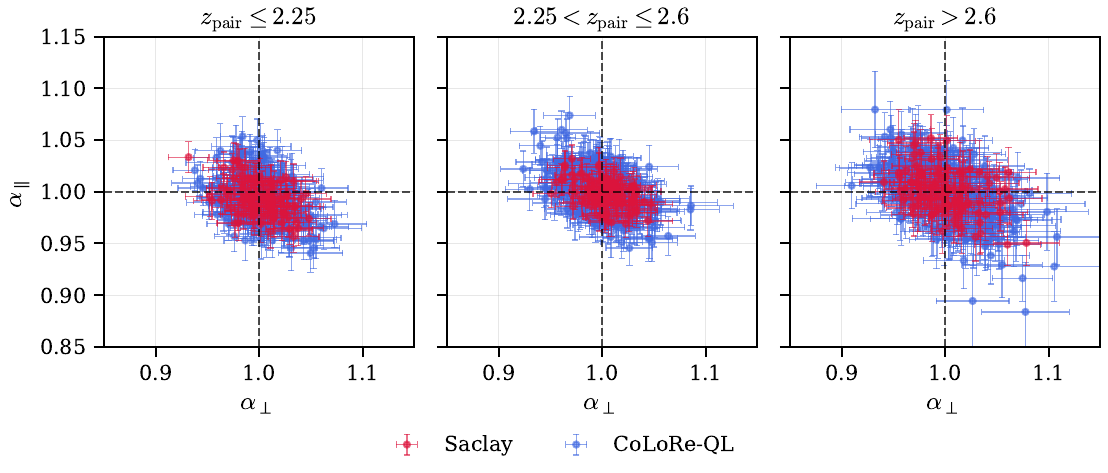}
    \caption{BAO scale parameters $\alpha_\parallel$ vs.\ $\alpha_\perp$ from fits to 100 \texttt{Saclay} mocks (red) and 300 \texttt{CoLoRe-QL} mocks (blue), in the three redshift bins. \textit{Top:} results from the stacked correlation functions. The black contour shows the $\sigma/3$ threshold computed from the baseline fit to the data in each redshift bin. \textit{Bottom:} scatter of best-fit values from individual fits. Dashed lines indicate the true input values $\alpha_\parallel = \alpha_\perp = 1$ in both panels.}
    \label{fig:mock_stacks_scatter}
\end{figure}

To assess potential biases, we fit the stacked correlation functions from each mock suite and compare the results to the true input values $\alpha_\parallel = \alpha_\perp = 1$. Following the criterion of \Cref{subsec:robustness}, we require that any deviation be smaller than $\sigma/3$. The top row of \Cref{fig:mock_stacks_scatter} shows that, in all cases, the recovered BAO parameters are consistent with the true values well within this threshold across all bins and both mock suites, indicating no detectable systematic bias at the level of precision of our measurements.

\begin{figure}[!tbp]
    \centering
    \includegraphics[width=\linewidth]{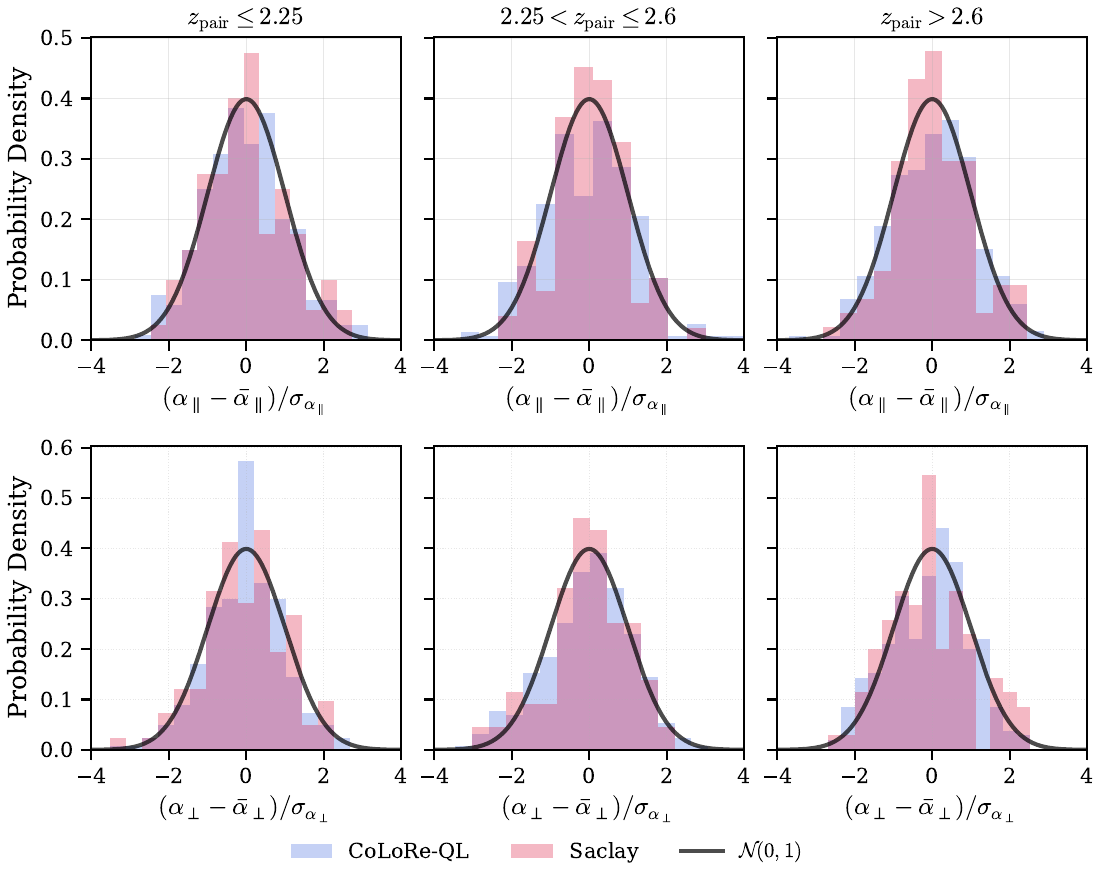}
    \caption{Pull distributions $(\alpha - \bar{\alpha})/\sigma_\alpha$ for $\alpha_\parallel$ (top row) and $\alpha_\perp$ (bottom row) from fits to 100 \texttt{Saclay} mocks (red) and 300 \texttt{CoLoRe-QL} mocks (blue), in the three redshift bins. The black curve shows the standard normal distribution $\mathcal{N}(0,1)$.}
    \label{fig:mock_pulls}
\end{figure}

We next examine the statistical behavior of the parameter estimates by fitting each realization individually. The bottom row of \Cref{fig:mock_stacks_scatter} shows the distribution of best-fit $(\alpha_\perp, \alpha_\parallel)$ values in each redshift bin. The measurements are centered on the true values with no significant outliers. The corresponding pull distributions $(\alpha - \bar{\alpha})/\sigma_\alpha$, shown in \Cref{fig:mock_pulls}, are consistent with a standard normal distribution $\mathcal{N}(0,1)$ for both mock suites and all redshift bins, indicating that the reported uncertainties are well calibrated and that the posterior distributions are approximately Gaussian.

\begin{figure}[!tbp]
    \centering
    \includegraphics[width=\linewidth]{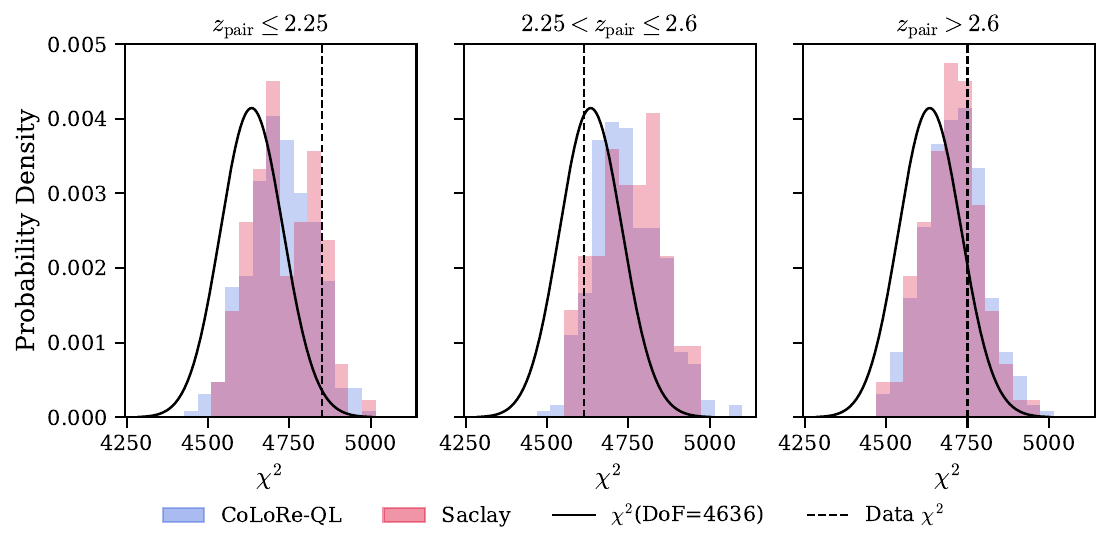}
    \caption{Distribution of best-fit $\chi^2$ values from fits to 300 \texttt{CoLoRe-QL} (blue) and 100 \texttt{Saclay} (red) mock realizations in the three redshift bins. The mock $\chi^2$ values are computed with the baseline (smoothed) covariance matrix per mock, whose effect on this distribution is examined in \Cref{fig:mock_chi2_cov}. The black curve shows the theoretical $\chi^2$ distribution for the corresponding number of degrees of freedom ($\mathrm{DoF} = 4636$), and the dashed vertical line indicates the best-fit $\chi^2$ from the data.}
    \label{fig:mock_chi2}
\end{figure}

Finally, we examine the distribution of best-fit $\chi^2$ values from individual fits. \Cref{fig:mock_chi2} compares the mock distributions to the theoretical expectation and to the best-fit $\chi^2$ values from the data. The mock distributions are shifted toward higher values relative to the theoretical prediction, a behavior previously observed in analyses of the complete dataset~\citep{KP6s6-Cuceu, Y3.lya-s1.Casas.2025}. This behavior is consistent with the $\chi^2$ values measured in the data (\Cref{subsec:goodnessoffit}); its origin is examined below.

To investigate the origin of this shift, we test the impact of the covariance matrix estimation. The \DESIDRIILya\ pipeline applies a smoothing step to the sub-sampling covariance to reduce noise, first introduced by \citep{Delubac:2015}, which may lead to a slight underestimation of the covariance amplitude. We repeat the fits to the \texttt{CoLoRe-QL} mocks using covariance matrices derived from the stack of all realizations, rescaled to a single realization, in both smoothed and unsmoothed forms.

\begin{figure}[!tbp]
    \centering
    \includegraphics[width=\linewidth]{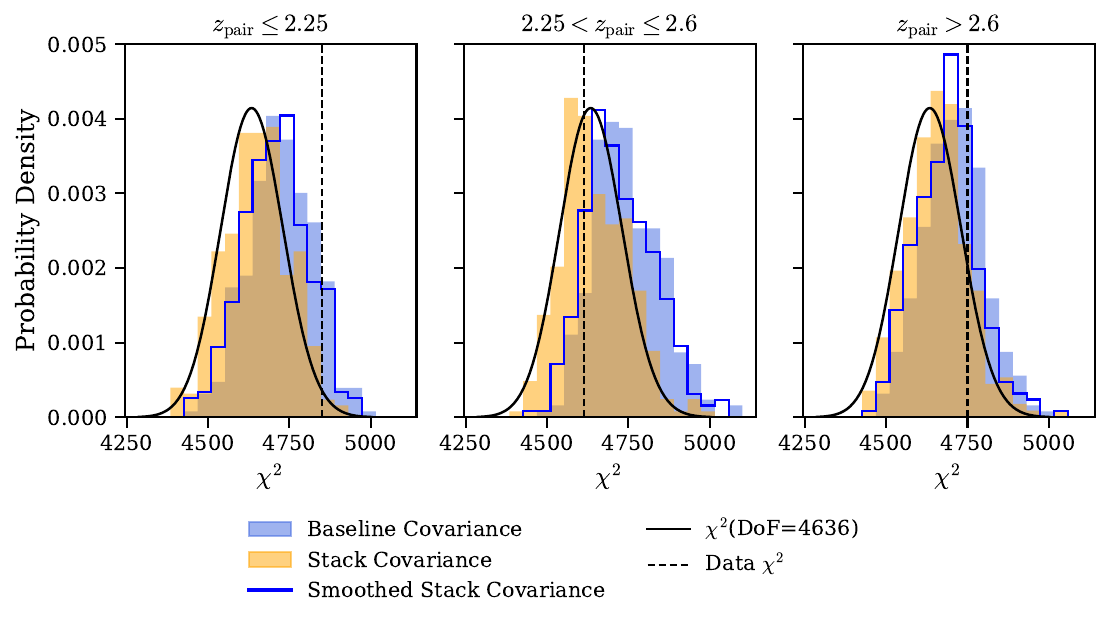}
    \caption{Distribution of best-fit $\chi^2$ values from fits to 300 \texttt{CoLoRe-QL} mock realizations in three redshift bins, obtained using three covariance matrices: the baseline per-mock smoothed covariance (blue filled histogram), the stack covariance rescaled to a single realization (orange filled histogram), and the smoothed stack covariance rescaled to a single realization (blue line histogram). The black curve shows the theoretical $\chi^2$ distribution for $\mathrm{DoF} = 4636$. The dashed vertical line indicates the best-fit $\chi^2$ from the data. }
    \label{fig:mock_chi2_cov}
\end{figure}

As shown in \Cref{fig:mock_chi2_cov}, using the unsmoothed stack covariance brings the $\chi^2$ distributions into significantly better agreement with the theoretical expectation, while the smoothed version reproduces the observed upward shift. The effect is small, with the median $\chi^2$ shifting by $1.1$--$2.3\%$ across the three redshift bins, and its impact on parameter uncertainties is negligible, corresponding to an underestimation of at most $\sim 1.15\%$.

This small effect does not produce a detectable change in the pull distributions, which remain consistent with $\mathcal{N}(0,1)$. A $1$--$2\%$ shift in $\chi^2$ translates into a comparably small change in the pull width, well below the statistical precision achievable with 400 realizations. Consistently, mock-by-mock comparisons show that the shifts in $\alpha_\parallel$ and $\alpha_\perp$ between covariance variants are negligible ($\lesssim 0.05\sigma$), indicating that the smoothing primarily affects the goodness-of-fit statistic rather than the parameter estimates.

Along with \Cref{subsec:robustness}, these results demonstrate that the BAO measurements reported in \Cref{tab:BAOparams} are robust. The mock analysis supports the interpretation of the observed $\chi^2$ values as consistent with expectations from synthetic datasets and shows that the pipeline delivers reliable parameter estimates with accurately calibrated uncertainties across all redshift bins.
\section{Cosmological Interpretation}\label{sec:inference}

Having established the robustness of our BAO measurements across three \lya\ redshift bins, we now turn to their cosmological interpretation.

We first convert the BAO scale parameters into physical distance ratios at the effective redshifts of each bin (\Cref{subsec:distances}). We then perform a direct, model-independent test of the matter-dominated expansion history enabled by our multi-redshift $D_H$ measurements (\Cref{subsec:eds}). Next, we quantify the additional information provided by the three-bin analysis relative to the single-bin \DESIDRIILya\ measurement (\Cref{subsec:constraining_power}), before finally deriving cosmological parameter constraints from DESI alone and in combination with external datasets (\Cref{subsec:cosmoparams}).

As mentioned in \Cref{sec:correlation}, the multi-redshift \lya\ measurements used in the cosmological fits throughout this section are derived from correlations extracted exclusively from the Ly$\alpha$(A) region, whereas the single-bin baseline measurements from \DESIDRIILya\ and \DESIDRIICosmo\ also incorporate information from the Ly$\alpha$(B) region. Consequently, differences between our results and the baseline may arise not only from the redshift binning strategy but also from the use of different datasets and from statistical fluctuations.

\subsection{Distance measurements}\label{subsec:distances}

Following the methodology of \citep{DESI.DR2.BAO.cosmo}, hereafter \DESIDRIICosmo, we convert the BAO scale parameters measured in the three redshift bins into physical distance ratios by multiplying each scale parameter by its value in the fiducial cosmology at the effective redshift of each bin. We report the transverse comoving distance $D_M/r_d$ and the Hubble distance $D_H/r_d$, together with the derived isotropic dilation parameter $D_V/r_d \equiv (z\,D_M^2\,D_H)^{1/3}/r_d$ and the anisotropic (Alcock-Paczyński) parameter $D_M/D_H$.

These measurements are summarized in \Cref{tab:distance_params}, together with the single-bin results from \DESIDRIILya\ at $z_{\rm eff}=2.33$ for comparison.

\begin{table}[t]
\centering
\begin{tabular}{lccc:cc}
    $z_{\rm eff}$ & $D_M/r_d$ & $D_H/r_d$ & $\rho_{D_M,D_H}$ & $D_V/r_d$ & $D_M/D_H$ \\
    \hline
    $2.13$ & $37.21\pm 0.82$ & $9.40\pm 0.20$ & $-0.469$ & $30.26\pm 0.39$ & $3.96\pm 0.15$ \\
    $2.40$ & $40.64\pm 0.99$ & $8.28\pm 0.17$ & $-0.467$ & $32.02\pm 0.46$ & $4.91\pm 0.19$ \\
    $2.81$ & $40.61^{+0.94}_{-1.1}$ & $7.22\pm 0.17$ & $-0.490$ & $32.22\pm 0.47$ & $5.63^{+0.22}_{-0.24}$ \\
    \hdashline
    $2.33$ & $38.99 \pm 0.53$ & $8.632 \pm 0.101$ & $-0.431$ & $31.27 \pm 0.26$  & $4.518 \pm 0.097$ \\
    \hline
\end{tabular}
\caption{Comoving distance measurements at the effective redshifts $z_{\rm eff}$ of the three \lya\ redshift bins. We report the transverse comoving distance $D_M/r_d$, the Hubble distance $D_H/r_d$, and their correlation coefficient $\rho_{D_M,D_H}$. We also include the isotropic dilation parameter $D_V/r_d \equiv (z\,D_M^2\,D_H)^{1/3}/r_d$ and the Alcock-Paczyński parameter $D_M/D_H$. For each redshift bin, the quoted values and $1\sigma$ uncertainties are obtained from the marginalized posterior distributions. A dashed line separates these results from the combined single-bin result at $z_{\rm eff}=2.33$, shown for comparison; for that row we quote the values reported by \DESIDRIILya, including both statistical and systematic uncertainties.}
\label{tab:distance_params}
\end{table}

\subsection{A test of Einstein-de Sitter expansion}\label{subsec:eds}

\begin{figure}
    \centering
    \includegraphics[width=0.9\linewidth]{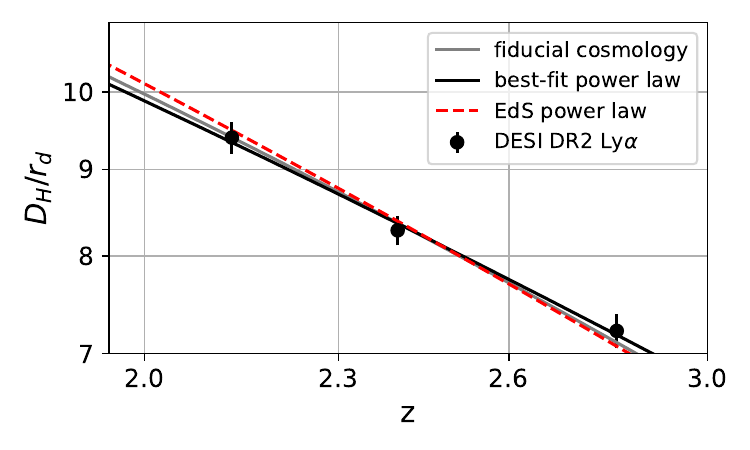}
    \caption{Longitudinal distance scale measurement as a function of redshift, in log-log scale (the horizontal axis being $\log(1+z)$). The grey curve is our fiducial $\Lambda$CDM model, the black line is a power-law fit to the data, and the red dashed line is an arbitrarily-normalized $(1+z)^{-3/2}$ power law.}
    \label{fig:eds_test}
\end{figure}

At high redshift $z \gtrsim 2$, the Universe is expected to be matter-dominated, so its expansion rate should be close to the Einstein-de Sitter regime~\cite{1932PNAS...18..213E},
\begin{equation}
    a(t) \propto t^{2/3} \qquad \mbox{or equivalently} \quad H(z) \propto (1+z)^{3/2}.
\end{equation}

This law is a direct consequence of Friedmann's equations when no additional component such as dark energy (or curvature) is added to the Universe's energy content. While this law is indirectly verified to high precision through joint fits of low-redshift probes and the CMB within a given cosmological model (e.g. comparing the matter-to-radiation density ratio, $\rho_m/\rho_\gamma$, at recombination with much less precise determinations of the same quantity today), a direct measurement of the expansion rate at these redshifts has not previously been available. The most closely related probe, cosmic chronometers~\citep{Jimenez2002}, provides cosmology-independent estimates of $H(z)$ but only reaches $z \sim 2$, and with substantially larger uncertainties~\citep{Moresco2015}; our measurement instead constrains the evolution of $H(z)$ over $2 \lesssim z \lesssim 3$.

Since the longitudinal BAO scale $D_H/r_d$ is directly proportional to $1/H(z)$, the multi-redshift measurement provides a first direct test of its redshift evolution, as illustrated in \Cref{fig:eds_test}. Assuming $H(z) \propto (1+z)^n$, a fit to the three redshift points yields $n = 1.34 \pm 0.16$, consistent with the matter-dominated (Einstein-de Sitter) expectation at the $1\sigma$ level.

Within the $\Lambda$CDM framework, dark energy still contributes at the $\sim 8\%$ level to the total energy density at $z=2$, so the $(1+z)^n$ scaling is only approximate: as shown in \Cref{fig:eds_test}, the $\Lambda$CDM prediction approaches the Einstein-de Sitter behavior at high redshift but deviates at $z \sim 2$. To quantify the discriminating power of the measurement, a synthetic data vector is constructed at the same redshifts and precision as the data, but assuming the fiducial cosmology (i.e. $\alpha_\parallel=1$ in all bins). Fitting this synthetic case yields $n = 1.42 \pm 0.16$.

The measured slope is therefore slightly closer to the $\Lambda$CDM expectation than to the Einstein-de Sitter case; however, the current statistical precision does not yet allow a significant distinction between the two. This constitutes a first, $12\%$-precision direct test of the matter-dominated expansion predicted by the Friedmann equations at these redshifts, and provides sensitivity to cosmological scenarios with non-standard evolution of the dark matter energy density.

\subsection{Constraining power of the redshift split}\label{subsec:constraining_power}

Splitting the \lya\ forest into three redshift bins reduces the statistical precision of each individual measurement while providing sensitivity to the redshift evolution of the distance scales. It is therefore not evident a priori whether this division leads to improved cosmological constraints. A direct comparison between the constraints obtained from the real single-bin and three-bin analyses would conflate changes in constraining power with statistical fluctuations in the measured central values. To isolate the former, we perform a forecast in which the distances are fixed to their fiducial values and only the covariance is varied, before deriving the cosmological parameter constraints presented in \Cref{subsec:cosmoparams}.

We build the forecast likelihood for two cases: one using the single-redshift \lya\ covariance of \DESIDRIILya, and one using the covariance built from the three \lya\ redshift bins introduced in this work. For all other tracers, the covariance is identical to that of the DESI cosmological parameter analysis~\DESIDRIICosmo.

Using these synthetic likelihoods, we perform the parameter inference with \texttt{Cobaya}~\citep{2019Cobaya,2021Cobaya} for three cosmological models: $\Lambda$CDM, its non-flat extension ($\Lambda$CDM$+\Omega_\mathrm{K}$), and a time-varying dark energy equation of state ($w_0w_a$CDM).

We quantify the improvement using the relative generalized figure of merit (FoM), $\mathrm{FoM}_r = 1/\sqrt{\det \mathrm{Cov}(p_1, p_2, \dots, p_i)}$, and compute the relative difference between the single-bin and three-bin cases. This difference is small for flat $\Lambda$CDM ($\sim1\%$) and moderate for $w_0w_a$CDM ($\sim8\%$), while for the non-flat model $\Lambda$CDM$+\Omega_\mathrm{K}$ the FoM improvement reaches ${\sim}15\%$ in the $(\Omega_\mathrm{m}, \Omega_\mathrm{K})$ plane.

\subsection{Cosmological parameter constraints}\label{subsec:cosmoparams}

Having assessed the constraining power of the three-bin \lya\ redshift split in \Cref{subsec:constraining_power}, we now derive cosmological constraints. We note that this analysis does not include the extensive systematic and data-split validation tests carried out in previous DESI \lya\ BAO analyses~\citep{DESI2024.IV.KP6, DESI.DR2.BAO.lya}. Moreover, the three-bin division adopted here was motivated by studying the redshift evolution of the model parameters while maintaining sufficient statistical precision, rather than by optimizing the cosmological constraining power. A dedicated optimization of the redshift binning strategy, together with the corresponding validation studies, is deferred to future work.

We consider three dataset configurations: the \lya\ forest alone, used as a consistency check; the full DESI DR2 BAO combination of galaxies, quasars, and the \lya\ forest, in which the forest enters either as a single redshift bin (the baseline analysis in \DESIDRIICosmo) or split into three bins (this work); and the latter in combination with external CMB and supernova data.

For the DESI-only configurations we adopt the same configuration as \DESIDRIICosmo, modifying only the \lya\ likelihood to incorporate the three redshift bins and their covariance; following Appendix~\ref{appendix:crossz_cov}, the cross-covariance between bins is treated as negligible. The marginalized constraints for all models and dataset combinations are summarized in \Cref{tab:cosmo_constraints}, alongside the baseline DESI DR2 results.

\begin{table*}
    \centering
    \resizebox{\linewidth}{!}{
    \begin{tabular}{lccccc}
    Model/Dataset & $\Omega_\mathrm{m}$ & $H_0 r_d$ [100 km s$^{-1}$] & $10^3\Omega_\mathrm{K}$ & $w_0$ & $w_a$ \\
    \hline
    \multicolumn{6}{l}{\textbf{$\Lambda$CDM}} \\
    DESI DR2 & $0.2975\pm 0.0086$ & $101.54\pm 0.73$ & --- & --- & --- \\
    DESI DR2 w/3 \lya\ bins & $0.3012\pm 0.0086$ & $101.33\pm 0.73$ & --- & --- & --- \\
    \hline
    \multicolumn{6}{l}{\textbf{$\Lambda$CDM$+\Omega_\mathrm{K}$}} \\
    DESI DR2 & $0.293\pm 0.012$ & $101.21\pm 0.90$ & $25\pm 41$ & --- & --- \\
    DESI DR2 w/3 \lya\ bins & $0.302\pm 0.010$ & $101.41\pm 0.89$ & $-5\pm 36$ & --- & --- \\
    DESI DR2 (1 \lya\ bin) + CMB-SPA + Dovekie & $0.3046\pm 0.0034$ & $100.80\pm 0.47$ & $2.5\pm 1.1$ & --- & --- \\
    DESI DR2 (3 \lya\ bins) + CMB-SPA + Dovekie & $0.3047\pm 0.0034$ & $100.82\pm 0.48$ & $2.8\pm 1.1$ & --- & --- \\
    \hline
    \multicolumn{6}{l}{\textbf{$w_0w_a$CDM}} \\
    DESI DR2 & $0.352^{+0.041}_{-0.018}$ & $94.7^{+2.2}_{-4.2}$ & --- & $-0.48^{+0.35}_{-0.17}$ & $< -1.34$ \\
    DESI DR2 w/3 \lya\ bins & $0.359^{+0.036}_{-0.018}$ & $94.4^{+2.1}_{-3.9}$ & --- & $-0.46^{+0.32}_{-0.16}$ & $< -1.52$ \\
    DESI DR2 (1 \lya\ bin) + CMB-SPA + Dovekie & $0.3128\pm 0.0054$ & $99.31\pm 0.82$ & --- & $-0.806\pm 0.054$ & $-0.73\pm 0.21$ \\
    DESI DR2 (3 \lya\ bins) + CMB-SPA + Dovekie & $0.3129\pm 0.0053$ & $99.28\pm 0.81$ & --- & $-0.807\pm 0.055$ & $-0.71^{+0.22}_{-0.19}$ \\
    \hline
    \end{tabular}
    }
    \caption{Cosmological parameter constraints from the full DESI DR2 BAO combination, with the \lya\ forest treated as a single bin (DESI DR2) or split into three bins (this work), and in combination with the external CMB-SPA and DES-Dovekie datasets. Values are marginalized posterior means with 68\% credible intervals, or 68\% upper limits where only one-sided constraints are available.}
    \label{tab:cosmo_constraints}
\end{table*}

\begin{figure}[t]
    \centering    \includegraphics[width=\linewidth]{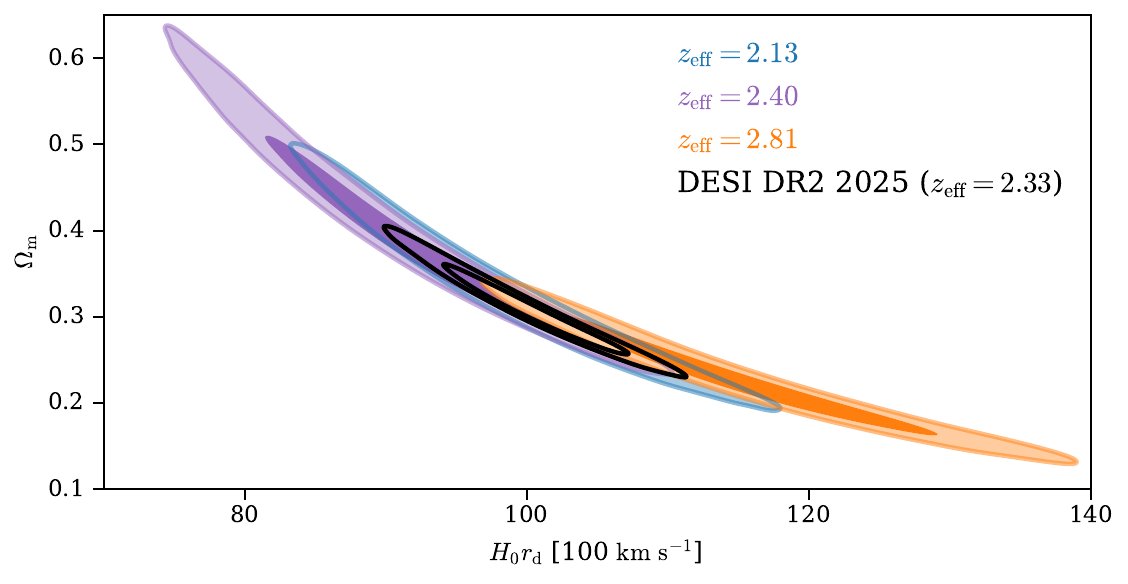}
    \caption{Flat $\Lambda$CDM constraints on $\Omega_\mathrm{m}$ and $H_0 r_d$ from the \lya\ data alone, in each of the three redshift bins independently, with the single-bin \lya\ measurement (black unfilled contour) overplotted for reference. The degeneracy direction does not change significantly between the effective redshifts.}
    \label{fig:onlyLya_posteriors}
\end{figure}

We first verify that the three \lya\ bins can be combined consistently within $\Lambda$CDM. \Cref{fig:onlyLya_posteriors} shows the flat $\Lambda$CDM constraints on $\Omega_\mathrm{m}$ and $H_0 r_d$ from the \lya\ data alone, in each bin independently, together with the single-bin \lya\ measurement from \DESIDRIILya\ for reference. The degeneracy direction does not change significantly over the parameter range where the contours overlap, and the constraints agree within $2\sigma$ across bins and with the DESI DR2 BAO measurements from galaxies and quasar clustering (see Figure 7 in \DESIDRIICosmo), which justifies combining the three-bin \lya\ measurement with them.

\begin{figure}[t]
    \centering
    \includegraphics[height=0.45\linewidth]{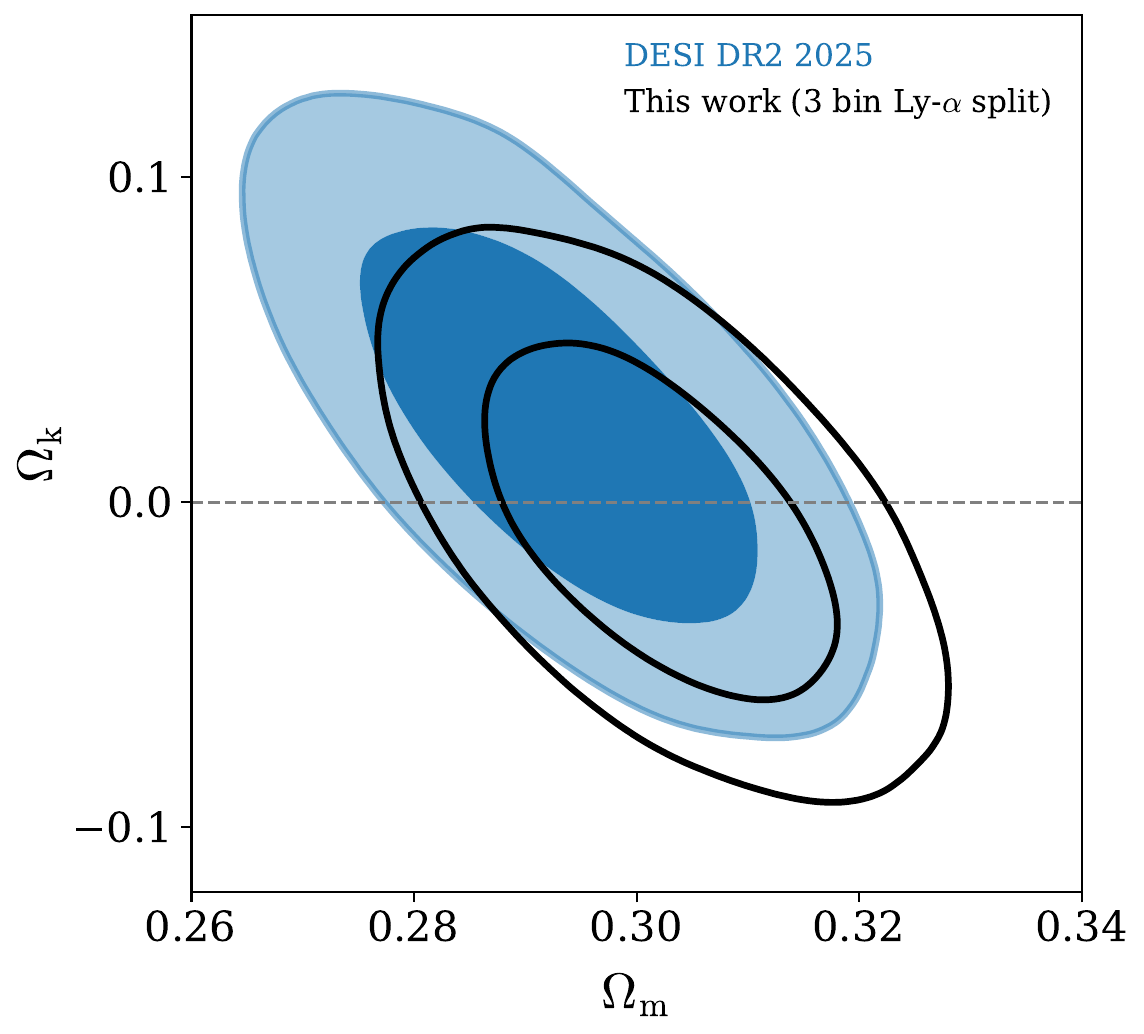}
    \includegraphics[height=0.45\linewidth]{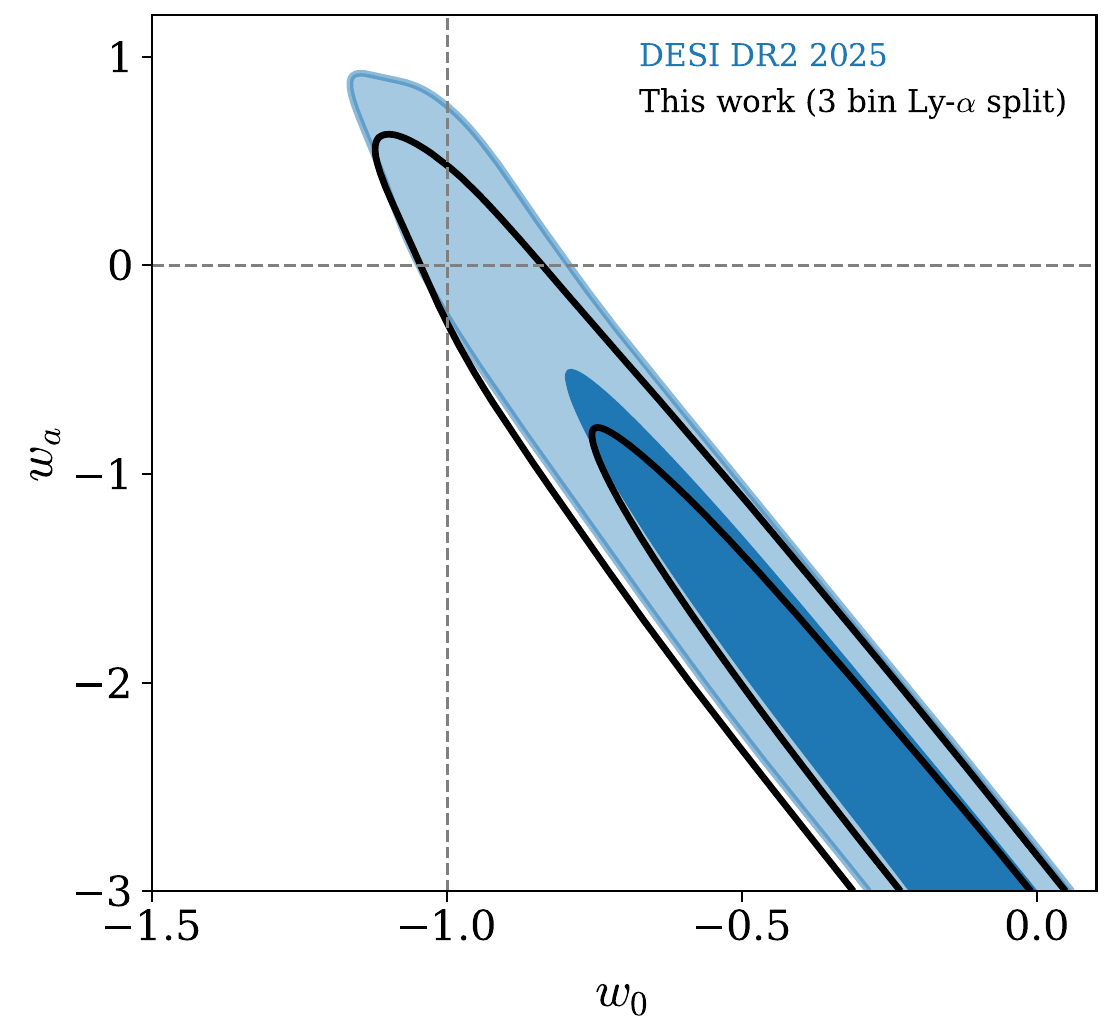}
    \caption{Constraints from the full DESI DR2 BAO combination with the \lya\ forest split into three bins (black contours), compared with the baseline single-bin analysis (blue)~\DESIDRIICosmo. Left: $\Omega_\mathrm{K}$ versus $\Omega_\mathrm{m}$ for the non-flat model $\Lambda$CDM$+\Omega_\mathrm{K}$. Right: $w_0$ versus $w_a$ for $w_0w_a$CDM.}
    \label{fig:data_desi_3bins}
\end{figure}

We then derive constraints from the full DESI BAO combination, comparing the baseline single-bin analysis with our three-bin split; the values of galaxies and quasars are identical to those used in \DESIDRIICosmo\ in both cases. \Cref{fig:data_desi_3bins} shows the resulting posteriors for the non-flat $\Lambda$CDM$+\Omega_\mathrm{K}$ and $w_0w_a$CDM models, where the redshift split is most relevant. 

For flat $\Lambda$CDM and $w_0w_a$CDM, we find no significant changes relative to the single-bin analysis: the posterior means shift slightly, but the two results remain consistent within $1\sigma$, with differences compatible with statistical fluctuations. The exception is the non-flat $\Lambda$CDM$+\Omega_\mathrm{K}$ model, as anticipated from the constraining-power analysis of \Cref{subsec:constraining_power}. In this case, the curvature constraint shifts from the single-bin value of $10^3\Omega_\mathrm{K} = 25 \pm 41$ to $10^3\Omega_\mathrm{K} = -5 \pm 36$, corresponding to a $\sim1\sigma$ shift in the central value and a $\sim12\%$ improvement in precision, in agreement with the forecast. Both measurements remain consistent with a flat Universe and with each other at the $1\sigma$ level.

Part of this shift may arise from our use of the Ly$\alpha$(A)-only dataset rather than from the redshift binning itself, since excluding the Ly$\alpha$(B) region degrades the precision of the full-sample \lya\ BAO measurement by approximately $10\%$. Nevertheless, the three-bin analysis yields a tighter curvature constraint than the single-bin baseline, indicating that the additional redshift information compensates for the loss of Ly$\alpha$(B) data. Future analyses incorporating the Ly$\alpha$(B) region are therefore expected to further improve the constraints.

Finally, we extend the analysis to combinations with external datasets, combining our BAO measurements with the recently recalibrated Dark Energy Survey supernova sample~\citep{Popovic:2025glk_recalibration,Dovekie_reanalysis_2025} (DES-Dovekie) and the CMB-SPA likelihood presented in~\citep{SPT-3G:2025bzu_CMB-SPA}, which combines Planck~\citep{Planck:2019nip_cmb-likelihoods,Planck:2018vyg_cosmological-parameters}, ACT~\citep{AtacamaCosmologyTelescope:2025blo_ACT-DR6}, and SPT primary-anisotropy data~\citep{SPT-3G:2025bzu_CMB-SPA}, together with the CMB lensing likelihood from~\citep{ACT:2025qjh_unified-cmb-lensing}.

\begin{figure}
    \centering
    \includegraphics[height=0.45\linewidth]{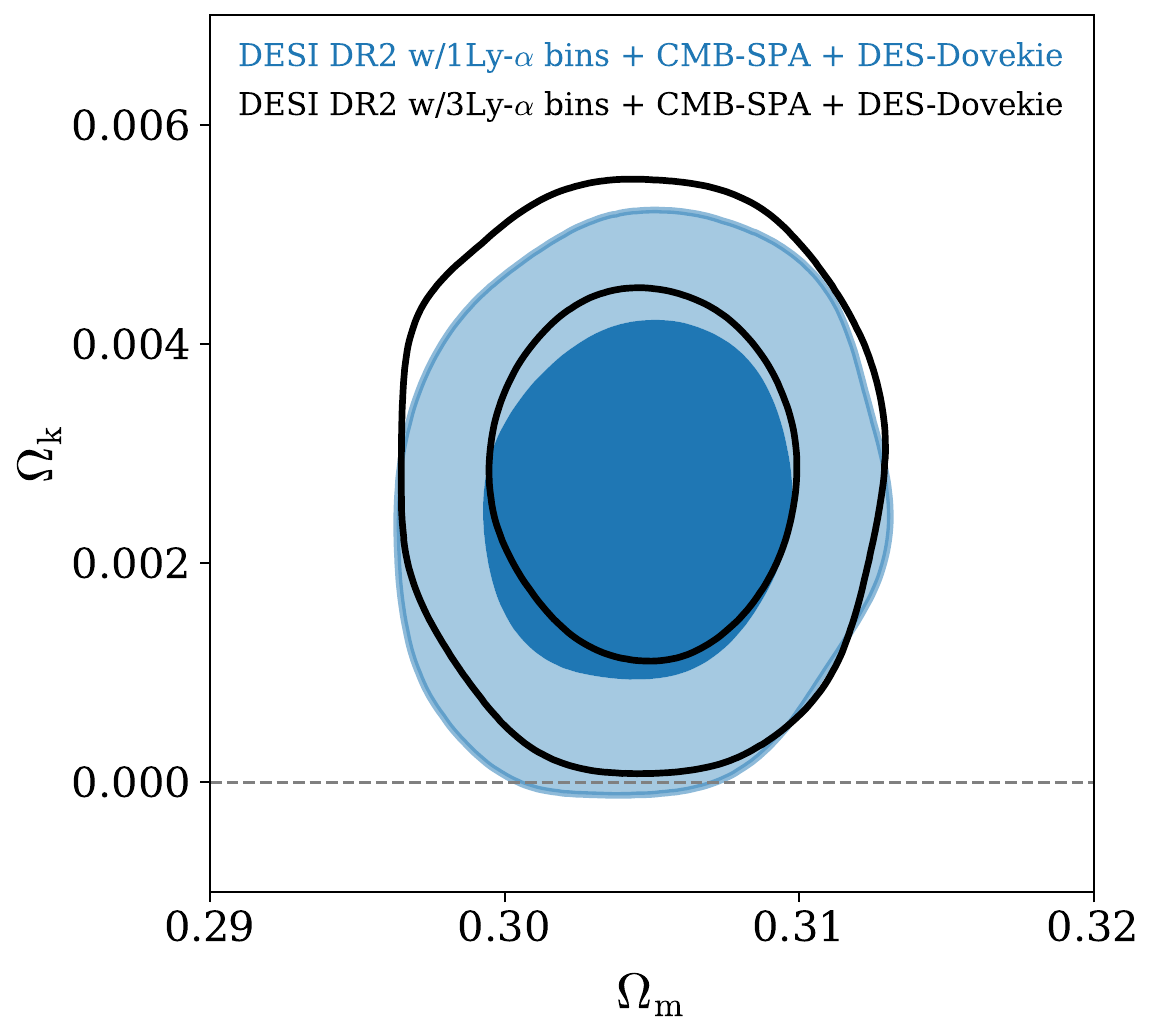}
    \includegraphics[height=0.45\linewidth]{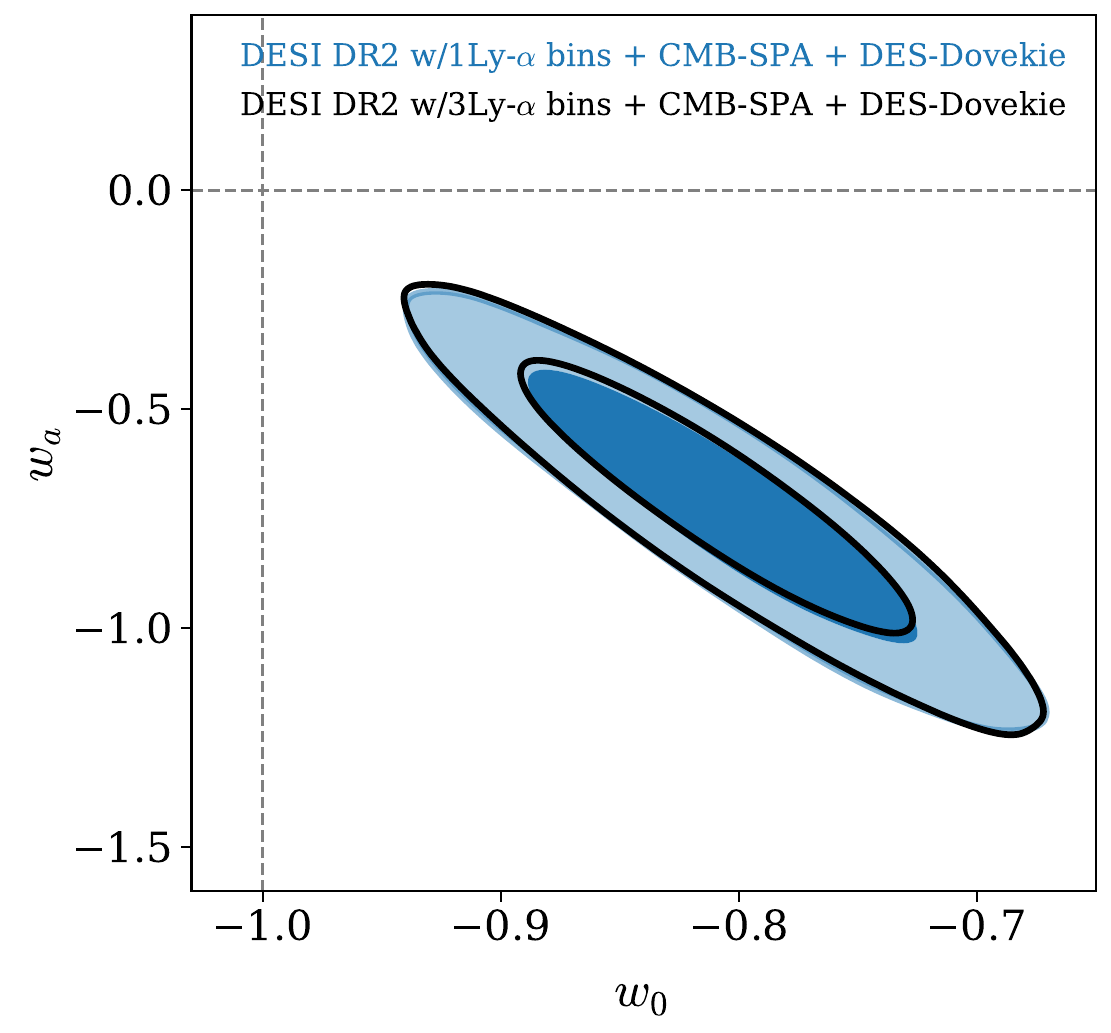}
    \caption{Constraints from DESI DR2 BAO with the \lya\ forest treated as a single bin (blue) or split into three bins (black contours), combined with the CMB-SPA likelihood and the DES-Dovekie supernova sample. Left: $\Omega_\mathrm{K}$ versus $\Omega_\mathrm{m}$ for $\Lambda$CDM$+\Omega_\mathrm{K}$. Right: $w_0$ versus $w_a$ for $w_0w_a$CDM.}
    \label{fig:desi_ext}
\end{figure}

\Cref{fig:desi_ext} shows the resulting constraints for the non-flat $\Lambda$CDM$+\Omega_\mathrm{K}$ and $w_0w_a$CDM models, comparing the single- and three-bin \lya\ analyses. In both cases, the two approaches remain consistent well within $1\sigma$, with parameter shifts below $\sim0.3\sigma$ (\Cref{tab:cosmo_constraints}). The uncertainties are essentially unchanged, indicating that although the multi-redshift \lya\ analysis provides additional information within the BAO dataset itself, its impact is subdominant once combined with the stronger constraining power of current CMB and supernova observations.

We note that the external dataset combination adopted here differs from that used in \DESIDRIICosmo. Consequently, this work is not intended as a direct comparison, and differences in the resulting posteriors may reflect these choices. A more detailed discussion of the impact of different external dataset combinations on cosmological parameter estimates is left to forthcoming DESI cosmological analyses.

\section{Summary and Conclusions}
\label{sec:conclusions}

We have presented BAO measurements from the DESI DR2 \lya\ forest split into three redshift bins, constituting the first multi-redshift \lya\ BAO analysis with DESI. Building on the DESI DR2 \lya\ data products and modeling framework, we introduced a pair-based redshift binning scheme in which each pixel-pixel or pixel-quasar pair is assigned to a bin according to its mean pair redshift $z_{\rm pair}$. This construction allows individual forests to contribute to multiple bins while ensuring that each pair contributes to exactly one bin. However, continuum-fitting distortions can introduce correlations across redshift bins; we quantify cross-bin correlations in Appendix~\ref{appendix:crossz_cov} and find them negligible at current precision. We adapted the correlation-function estimators, distortion matrices, and metal templates to this binning scheme, and jointly fitted the \lya\ auto- and cross-correlations in three redshift intervals defined by $z_{\rm pair}\leq2.25$, $2.25<z_{\rm pair}\leq2.6$, and $z_{\rm pair}>2.6$, corresponding to effective redshifts $z_{\rm eff}=2.13$, $2.40$, and $2.81$. Two results stand out: a direct measurement of the matter-dominated expansion rate over $2\lesssim z\lesssim3$, and a self-consistent determination of the redshift evolution of the \lya\ forest and quasar bias parameters within a single analysis.

The combined fits yield isotropic BAO constraints with a precision of approximately $1.1$--$1.2\%$ per bin (\Cref{tab:BAOparams}). The intermediate and highest-redshift bins exhibit acceptable goodness of fit, with ${\rm PTE}=0.59$ and $0.12$, respectively, while the lowest-redshift bin shows a lower value ($\mathrm{PTE}=0.01$). We trace this feature to localized systematics on small scales and near the line of sight (where continuum-fitting distortion, residual quasar redshift errors, and unmodeled small-scale contaminants may all contribute), combined with a covariance-smoothing effect. We find that restricting the fitting range, limiting the $\mu$ range, or accounting for the covariance shift each restores an acceptable PTE without appreciably shifting the recovered BAO scale. More generally, our measurements are stable against a broad set of robustness tests spanning fitting ranges, modeling choices, and prior variations, and the auto- and cross-correlation measurements are mutually consistent, apart from a marginal offset in the intermediate redshift bin that is compatible with a statistical fluctuation.

We validated the full analysis pipeline using $300$ \texttt{CoLoRe-QL} and $100$ \texttt{Saclay} synthetic realizations, processed with the same pair-based binning scheme. Fits to stacked correlation functions recover the unbiased values well within the $\sigma/3$ threshold in all bins and both mock suites. The pull distributions from individual realizations are consistent with a unit normal distribution, indicating well-calibrated statistical uncertainties. The mock $\chi^2$ distributions are systematically shifted toward higher values relative to the nominal expectation, reproducing the behavior observed in the data; we show that this arises from smoothing in the sub-sampling covariance estimation, which shifts the median $\chi^2$ by $1.1$--$2.3\%$ across bins while affecting parameter uncertainties at the level of $\lesssim1.15\%$.

Beyond the BAO scale, the joint multi-redshift analysis directly constrains the redshift evolution of \lya\ forest and quasar bias parameters within a single self-consistent framework. The effective \lya\ bias becomes increasingly negative with redshift ($\gamma_\alpha=3.05\pm0.16$), consistent with the commonly adopted scaling $\gamma_\alpha=2.9$. The RSD parameter $\beta'_\alpha$ decreases with redshift ($\gamma_\beta=-0.97\pm0.26$), following a trend also observed in hydrodynamical simulations~\citep{Arinyo2015,Chabanier2024}. A dedicated study of DLA masking effects will be required to fully assess the robustness of this evolution. The quasar bias increases as $\gamma_Q=1.56\pm0.23$, in agreement with standard expectations and independent DESI quasar clustering measurements, providing a consistency check across tracers.

Translating the BAO parameters into distance measurements (\Cref{tab:distance_params}), the three bins constrain the transverse comoving distance $D_M/r_d$ and the Hubble distance $D_H/r_d$ at $z_{\rm eff}=2.13$, $2.40$, and $2.81$. The transverse distance $D_M/r_d$ spans $37.21\pm0.82$ to $\simeq40.6$, providing an integral probe of the comoving distance, while the anisotropic ratio $D_M/D_H$ grows monotonically from $3.96\pm0.15$ to $5.63^{+0.22}_{-0.24}$ across the bins. The radial measurement $D_H/r_d$ decreases from $9.40\pm0.20$ to $7.22\pm0.17$, directly probing the local evolution of $H(z)$ in the matter-dominated epoch.

Fitting a power-law model $H(z)\propto(1+z)^n$ to the three radial measurements yields $n=1.34\pm0.16$, a $12\%$ measurement of the logarithmic slope, consistent with the Einstein-de Sitter expectation $n=3/2$ at the $1\sigma$ level. A synthetic realization drawn from our fiducial $\Lambda$CDM\ cosmology yields $n=1.42\pm0.16$, indicating slightly better agreement with $\Lambda$CDM\ than with a pure Einstein-de Sitter model, although current uncertainties are not yet sufficient to distinguish between them. This constitutes a first direct test of the matter-dominated expansion predicted by the Friedmann equations in this redshift regime, consistent with the standard picture.

We verified that the three redshift bins are consistent with being statistically independent, with cross-bin correlations consistent with zero (Appendix~\ref{appendix:crossz_cov}), allowing them to be combined straightforwardly with the other DESI DR2 BAO tracers. Doing so, we find that cosmological constraints are largely insensitive to the binning scheme for flat $\Lambda$CDM\ and $w_0w_a$CDM models, with changes in the generalized figure of merit of $\sim1\%$ and $\sim8\%$, respectively. The most significant impact appears in non-flat $\Lambda$CDM$+\Omega_\mathrm{K}$, where the three-bin analysis improves the figure of merit in the $(\Omega_\mathrm{m},\Omega_\mathrm{K})$ plane by ${\sim}15\%$ and tightens the marginalized curvature constraint by ${\sim}12\%$, shifting it from $10^3\Omega_\mathrm{K}=25\pm41$ to $10^3\Omega_\mathrm{K}=-5\pm36$; both results remain fully consistent with spatial flatness and with each other at the $1\sigma$ level. When combined with the external CMB-SPA and DES-Dovekie datasets, the single- and three-bin analyses remain in close agreement, with parameter shifts below $\sim0.3\sigma$ and essentially unchanged uncertainties, indicating that the redshift splitting does not materially affect the joint cosmological constraints.

Looking forward, several improvements to the present analysis are within reach. As noted in \Cref{subsec:cosmoparams}, a dedicated optimization of the redshift binning, together with a complete validation pipeline consistent with previous DESI \lya\ BAO analyses~\citep{DESI2024.IV.KP6, DESI.DR2.BAO.lya}, constitutes a natural next step. Within such a framework, a blinding procedure tailored to multi-redshift measurements would provide an additional safeguard against confirmation bias, extending the blinding approaches implemented in previous DESI \lya\ BAO analyses.

Beyond validation aspects, improvements to the statistical power of the measurement are also possible. In this work, we have restricted the measurement to \lya\ region~A correlations; extending the analysis to include region~B is expected to improve BAO precision by $\sim10\%$, at the cost of a more complex modeling of the cross-redshift covariance (Appendix~\ref{appendix:crossz_cov}). Additional information can be recovered through a full-shape analysis of the redshift-split correlations, which exploits the broadband signal beyond the BAO peak and can significantly reduce uncertainties in the Alcock-Paczyński parameter~\citep{Cuceu:2021FS, Cuceu:2023eBOSS, Cuceu:2025DR1}.

At the survey level, multi-redshift \lya\ analyses will become increasingly relevant for upcoming experiments such as DESI Run~2~\citep{DESI-II:2022}, Spec-S5~\citep{Spec-S5:2025uom}, WST~\citep{WST:2024zvm}, and MUST~\citep{Zhao:2024MUST}. In particular, combining these measurements with forests in Lyman-break galaxies will extend the accessible redshift range to $z \gtrsim 2.5$~\citep{Herrera:LBGs}, enabling finer redshift resolution and improved sensitivity to the high-redshift expansion history. At this level of precision, multi-redshift \lya\ analyses become a uniquely powerful probe of the expansion history deep in the matter-dominated era.
\section*{Data Availability}
The data used in this analysis will be made public with DESI Data Release 2 (details at \url{https://data.desi.lbl.gov/doc/releases/}). The data points corresponding to the figures in this paper will be made available upon publication in the Zenodo repository at \url{https://doi.org/10.5281/zenodo.22710550}, in compliance with the DESI Data Management Plan.
\section*{Acknowledgements}
HKHA and EA acknowledge the funding of the French Agence Nationale de la Recherche (ANR) under grant ANR-22-CE31-0009 (HZ3DMAP project) and grant ANR-22-CE92-0037 (DESILya project). AXGM acknowledges the C\'atedra Moshinsky 2024 funding. AXGM and EP acknowledge the funding from SECIHTI under grants CBF-2025-G-1327 and the PhD program, the DAIP P0409.026.301 grant and the UGData Lab. CGQ acknowledges the support provided by NASA through the NASA Hubble Fellowship grant HST-HF2-51554.001-A awarded by the Space Telescope Science Institute, which is operated by the Association of Universities for Research in Astronomy, Inc., for NASA, under contract NAS5-26555.

This material is based upon work supported by the U.S. Department of Energy (DOE), Office of Science, Office of High-Energy Physics, under Contract No. DE–AC02–05CH11231, and by the National Energy Research Scientific Computing Center, a DOE Office of Science User Facility under the same contract. Additional support for DESI was provided by the U.S. National Science Foundation (NSF), Division of Astronomical Sciences under Contract No. AST-0950945 to the NSF’s National Optical-Infrared Astronomy Research Laboratory; the Science and Technology Facilities Council of the United Kingdom; the Gordon and Betty Moore Foundation; the Heising-Simons Foundation; the French Alternative Energies and Atomic Energy Commission (CEA); the Secretariat of Science, Humanities, Technology and Innovation (SECIHTI) of Mexico; the Ministry of Science, Innovation and Universities of Spain (MICIU/AEI/10.13039/501100011033), and by the DESI Member Institutions: \url{https://www.desi.lbl.gov/collaborating-institutions}. Any opinions, findings, and conclusions or recommendations expressed in this material are those of the author(s) and do not necessarily reflect the views of the U. S. National Science Foundation, the U. S. Department of Energy, or any of the listed funding agencies.

The authors are honored to be permitted to conduct scientific research on I'oligam Du'ag (Kitt Peak), a mountain with particular significance to the Tohono O’odham Nation.

\bibliographystyle{JHEP.bst}
\bibliography{references,DESI_supporting_papers}

\appendix 
\counterwithin{figure}{section}
\counterwithin{table}{section}
\section{Covariance Across Redshift Bins}\label{appendix:crossz_cov}

The three redshift bins defined in \Cref{sec:analysis} are constructed from the same set of quasar spectra and tracers. Individual forest pixels and quasars can therefore contribute to multiple bins through distinct pixel–pixel or pixel–quasar pairs assigned according to $z_{\rm pair}$. Furthermore, as discussed in \Cref{sec:correlation,subsec:dmat}, continuum-fitting distortions can induce correlations across redshift bins. As a result, the measured correlation functions are not strictly independent, and these effects may propagate into correlations between the corresponding BAO parameter estimates.

To assess the significance of possible cross-bin correlations, we follow an approach analogous to that used for the DESI-SDSS cross-survey covariance in Appendix~F of \citep{DESI2024.IV.KP6}, based on Monte Carlo realizations. A full joint analysis of all three bins is computationally prohibitive, so we estimate the cross-bin covariance pairwise.

For each pair of redshift bins $(Z_i,Z_j)$, we generate $N_\mathrm{MC} = 6144$ Monte Carlo realizations of the correlation functions using their joint covariance matrix. This joint covariance is obtained from the same sub-sampling estimator used for the per-bin covariances in \DESIDRIILya, followed by the same smoothing procedure, and includes the cross-bin covariance as off-diagonal blocks. Each realization is fitted independently to obtain $(\alpha_\parallel, \alpha_\perp)$, and the Pearson correlation coefficients between bins are computed. Uncertainties on these coefficients are estimated via bootstrap resampling.

The resulting $6\times6$ correlation matrix of the BAO parameters is
\begin{equation}
\resizebox{0.9\textwidth}{!}{$
\begin{pmatrix}
1 & \rho(\alpha_\parallel^1,\alpha_\perp^1) &
    \rho(\alpha_\parallel^1,\alpha_\parallel^2) &
    \rho(\alpha_\parallel^1,\alpha_\perp^2) &
    \rho(\alpha_\parallel^1,\alpha_\parallel^3) &
    \rho(\alpha_\parallel^1,\alpha_\perp^3) \\[4pt]
\rho(\alpha_\perp^1,\alpha_\parallel^1) & 1 &
    \rho(\alpha_\perp^1,\alpha_\parallel^2) &
    \rho(\alpha_\perp^1,\alpha_\perp^2) &
    \rho(\alpha_\perp^1,\alpha_\parallel^3) &
    \rho(\alpha_\perp^1,\alpha_\perp^3) \\[4pt]
\rho(\alpha_\parallel^2,\alpha_\parallel^1) &
    \rho(\alpha_\parallel^2,\alpha_\perp^1) & 1 &
    \rho(\alpha_\parallel^2,\alpha_\perp^2) &
    \rho(\alpha_\parallel^2,\alpha_\parallel^3) &
    \rho(\alpha_\parallel^2,\alpha_\perp^3) \\[4pt]
\rho(\alpha_\perp^2,\alpha_\parallel^1) &
    \rho(\alpha_\perp^2,\alpha_\perp^1) &
    \rho(\alpha_\perp^2,\alpha_\parallel^2) & 1 &
    \rho(\alpha_\perp^2,\alpha_\parallel^3) &
    \rho(\alpha_\perp^2,\alpha_\perp^3) \\[4pt]
\rho(\alpha_\parallel^3,\alpha_\parallel^1) &
    \rho(\alpha_\parallel^3,\alpha_\perp^1) &
    \rho(\alpha_\parallel^3,\alpha_\parallel^2) &
    \rho(\alpha_\parallel^3,\alpha_\perp^2) & 1 &
    \rho(\alpha_\parallel^3,\alpha_\perp^3) \\[4pt]
\rho(\alpha_\perp^3,\alpha_\parallel^1) &
    \rho(\alpha_\perp^3,\alpha_\perp^1) &
    \rho(\alpha_\perp^3,\alpha_\parallel^2) &
    \rho(\alpha_\perp^3,\alpha_\perp^2) &
    \rho(\alpha_\perp^3,\alpha_\parallel^3) & 1
\end{pmatrix}
=
\begin{pmatrix}
 1.00  & -0.50 &  0.00 & -0.02 &  0.01 & -0.02 \\[4pt]
-0.50  &  1.00 & -0.01 &  0.02 & -0.01 &  0.02 \\[4pt]
 0.00  & -0.01 &  1.00 & -0.51 &  0.00 & -0.02 \\[4pt]
-0.02  &  0.02 & -0.51 &  1.00 & -0.01 &  0.02 \\[4pt]
 0.01  & -0.01 &  0.00 & -0.01 &  1.00 & -0.52 \\[4pt]
-0.02  &  0.02 & -0.02 &  0.02 & -0.52 &  1.00
\end{pmatrix}$},
\label{eq:crossz_corrmat}
\end{equation}
where superscripts $1$, $2$, and $3$ denote the bins $z_\mathrm{pair} \leq 2.25$, $2.25 < z_\mathrm{pair} \leq 2.6$, and $z_\mathrm{pair} > 2.6$, respectively, and the parameter ordering within each bin is $(\alpha_\parallel, \alpha_\perp)$.

All cross-bin correlations are consistent with zero, with values in the range $[-0.02, +0.02]$, indicating that any correlation induced by continuum-fitting distortions or shared individual tracers is below the statistical sensitivity of this analysis. The statistical uncertainty on the correlation coefficients is $\sigma_\rho \simeq 0.013$, implying that correlations with $|\rho| \gtrsim 0.04$ would be required for a $3\sigma$ detection.

\begin{figure}[!tbp]
    \centering
    \includegraphics[width=\textwidth]{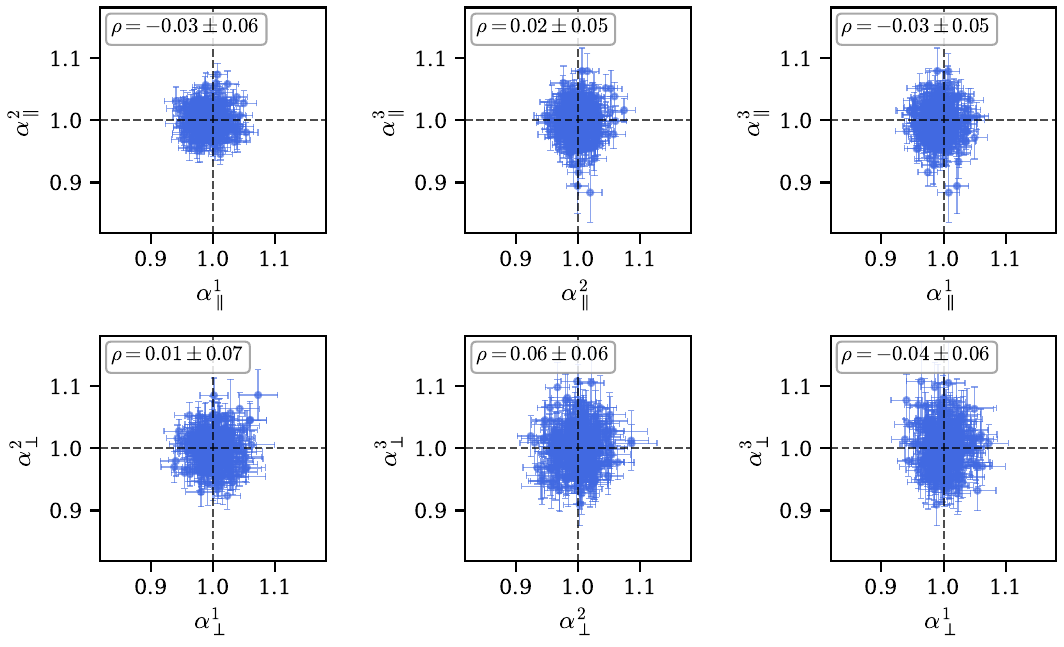}
    \caption{Scatter of the best-fit BAO scale parameters across the 300 \texttt{CoLoRe-QL} mock realizations, showing the same-parameter cross-bin correlations ($\alpha_\parallel$ with $\alpha_\parallel$, top row, and $\alpha_\perp$ with $\alpha_\perp$, bottom row) for each pair of redshift bins. The superscripts $1$, $2$, and $3$ correspond to the bins $z_{\rm pair}\leq2.25$, $2.25<z_{\rm pair}\leq2.6$, and $z_{\rm pair}>2.6$, as in \Cref{eq:crossz_corrmat}. The dashed lines indicate the fiducial value of unity. The Pearson correlation coefficient $\rho$ is reported in each panel together with its bootstrap uncertainty.}
    \label{fig:crossz_scatter}
\end{figure}

We validate this result using the 300 \texttt{CoLoRe-QL} mock realizations described in \Cref{sec:validation}, estimating cross-bin correlations directly from the scatter of the best-fit BAO parameters across realizations. \Cref{fig:crossz_scatter} shows the same-parameter cross-bin correlations ($\alpha_\parallel$ with $\alpha_\parallel$ and $\alpha_\perp$ with $\alpha_\perp$). All measured correlations are consistent with zero within their bootstrap uncertainties, the largest being $\rho = 0.06 \pm 0.06$ for $\alpha_\perp$ between the second and third redshift bins. The mixed $\alpha_\parallel$--$\alpha_\perp$ cross-bin correlations are likewise consistent with zero. The finite number of mock realizations limits the sensitivity of this test to $|\rho| \lesssim 0.06$, consistent with the tighter $[-0.02, +0.02]$ interval obtained from the Monte Carlo realizations.

The within-bin $\alpha_\parallel$–$\alpha_\perp$ correlations (the $2\times2$ diagonal blocks) are $\rho \simeq -0.50$ to $-0.52$ in the Monte Carlo analysis. Since these are derived from the covariance-based description of the data, they may not fully capture all parameter degeneracies. We therefore adopt the correlations measured directly from the posterior distributions in \Cref{tab:BAOparams}: $\rho(\alpha_\parallel^1, \alpha_\perp^1) = -0.469$, $\rho(\alpha_\parallel^2, \alpha_\perp^2) = -0.467$, and $\rho(\alpha_\parallel^3, \alpha_\perp^3) = -0.490$.

The final $6\times6$ covariance matrix used for cosmological interpretation is constructed from these within-bin correlations and the per-bin BAO parameter uncertainties from \Cref{tab:BAOparams}, with all cross-bin entries set to zero.

A similar conclusion holds for the nuisance parameters: their cross-bin correlations are consistent with Monte Carlo noise and show no coherent trend. A mild exception is the quasar bias parameter $b_Q$, for which we find a weak anti-correlation between the first and second bins ($\rho \sim -0.08$) and a smaller correlation between the second and third bins ($\rho \sim -0.04$), while no significant correlation is observed between the first and third bins.
\section{Nuisance Parameter Constraints}\label{appendix:nuisance}
In this appendix we report the complete set of nuisance-parameter marginalized constraints (\Cref{tab:Allparams}) from the combined fits described in \Cref{sec:results}. All constraints assume a fixed Planck 2018 $\Lambda$CDM fiducial cosmology~\citep{Planck:2018vyg_cosmological-parameters}; parameters such as biases and RSD coefficients are sensitive to this assumption and may shift if it is relaxed.

The redshift evolution of the primary \lya\ forest and quasar clustering parameters, namely the effective bias $b_\alpha'$, the RSD parameter $\beta_\alpha'$, and the quasar bias $b_Q$, was presented in \Cref{subsec:nuisance_evol}. Here we discuss the remaining nuisance parameters: the quasar redshift systematics, the transverse proximity effect, the metal and HCD parameters, and the correlated noise.

\begin{table}[t]
\centering
\resizebox{\textwidth}{!}{
\begin{tabular}{l|c|lll}
Parameter  & Prior & $z_{\rm pair} \leq 2.25$ & 
    $2.25 < z_{\rm pair} \leq 2.6$ & $z_{\rm pair} > 2.6$  \\
\hline
$\alpha_{\parallel}$ & $\mathcal{U}[0.01, 2.00]$ & $0.9997\pm0.022$ & $0.990\pm0.020$ & $1.016\pm0.024$ \\
$\alpha_{\perp}$ & $\mathcal{U}[0.01, 2.00]$ & $0.995\pm0.022$ & $1.021\pm0.025$ & $0.946^{+0.022}_{-0.025}$ \\ 
$b_{\alpha}$ & $\mathcal{U}[-2.00, 0.00]$ & $-0.0703^{+0.012}_{-0.0093}$ & $-0.1428^{+0.0093}_{-0.015}$ & $-0.2286^{+0.0052}_{-0.011}$\\
$\beta_{\alpha}$ & $\mathcal{U}[0.00, 5.00]$ & $2.25^{+0.26}_{-0.35}$ & $1.469^{+0.071}_{-0.14}$ & $1.208^{+0.047}_{-0.069}$\\
$b_{Q}$ & $\mathcal{U}[0.00, 6.00]$ & $3.206\pm 0.088$ & $3.731\pm 0.098$ & $4.35^{+0.15}_{-0.17}$\\
$\sigma_z\;[h^{-1}\mathrm{Mpc}]$ & $\mathcal{U}[0.00, 15.00]$ & $2.99^{+1.2}_{-0.94}$ & $3.42^{+1.1}_{-0.76}$ & $2.3^{+1.1}_{-1.8}$\\
$\Delta r_{\parallel}\;[h^{-1}\mathrm{Mpc}]$ & $\mathcal{N}(0.0, 1.0)$ & $1.06\pm 0.29$ & $0.61\pm 0.29$ & $0.39\pm 0.36$\\
$\xi_0^{\rm TP}$ & $\mathcal{U}[0.00, 2.00]$ & $0.310\pm 0.064$ & $0.489\pm 0.082$ & $0.86\pm 0.18$\\
$b_{\rm HCD}$ & $\mathcal{U}[-0.20, 0.00]$ & $-0.061^{+0.010}_{-0.014}$ & $-0.024^{+0.019}_{-0.011}$ & $-0.0102^{+0.010}_{-0.0023}$\\
$\beta_{\rm HCD}$ & $\mathcal{N}(0.500, 0.090)$ & $0.522\pm 0.088$ & $0.505\pm 0.089$ & $0.492\pm 0.092$\\
$L_{\rm HCD}\;[h^{-1}\mathrm{Mpc}]$ & $\mathcal{N}(5.0, 1.0)$ & $5.62\pm 0.83$ & $4.8\pm 1.0$ & $4.8\pm 1.0$\\
$10^3\,b_{\rm SiII(1190)}$ & $\mathcal{U}[-500, 0]$ & $-2.68\pm 0.63$ & $-4.75\pm 0.63$ & $-4.9\pm 1.1$\\
$10^3\,b_{\rm SiII(1193)}$ & $\mathcal{U}[-500, 0]$ & $-4.11\pm 0.65$ & $-2.87\pm 0.61$ & $-2.07^{+1.1}_{-0.91}$\\
$10^3\,b_{\rm SiIII(1207)}$ & $\mathcal{U}[-500, 0]$ & $-5.7\pm 2.0$ & $-7.6\pm 2.4$ & $-1.63^{+1.6}_{-0.45}$\\
$10^3\,b_{\rm SiII(1260)}$ & $\mathcal{U}[-500, 0]$ & $-4.35\pm 0.61$ & $-3.90\pm 0.60$ & $-5.9\pm 1.0$\\
$10^3\,b_{\rm CIV(eff)}$ & $\mathcal{N}(-19.0, 5.0)$ & $-18.6\pm 4.9$ & $-19.2\pm 5.0$ & $-19.1\pm 5.0$\\
$10^4\,a_{\rm noise}$ & $\mathcal{U}[0, 100]$ & $2.42\pm 0.24$ & $1.83\pm 0.27$ & $0.93^{+0.42}_{-0.55}$\\
\hdashline
$b_{\alpha}'$ & -- & $-0.1317\pm 0.0035$ & $-0.1670^{+0.0045}_{-0.0041}$ & $-0.2388\pm 0.0045$\\
$\beta_{\alpha}'$ & -- & $1.423\pm 0.048$ & $1.322\pm 0.045$ & $1.176\pm 0.046$\\
\hline
$z_{\rm eff}$ & -- & $2.13$ & $2.40$ & $2.81$ \\
$\chi_{\rm min}^2$ & -- & $4849.24$ & $4612.68$ & $4750.15$ \\
DoF & -- & $4653-17$ & $4653-17$ & $4653-17$ \\
PTE & -- & $0.01$ & $0.59$ & $0.12$ \\
\hline
\end{tabular}}
\caption{Marginalized posterior constraints on all fitted parameters in the three redshift bins. Quoted uncertainties correspond to $1\sigma$ credible intervals. The effective \lya\ bias $b_\alpha'$ and $\beta_\alpha'$ (separated by a dashed rule) are derived quantities defined in \Cref{eq:beff,eq:betaeff}. Parameters below the solid rule are goodness-of-fit statistics. All nuisance parameter constraints assume a fixed Planck 2018 $\Lambda$CDM fiducial cosmology~\citep{Planck:2018vyg_cosmological-parameters}.}
\label{tab:Allparams}
\end{table}

\subsection{Quasar redshift systematics and the proximity effect}

The cross-correlation constrains two parameters describing line-of-sight systematics associated with quasar redshift uncertainties: the statistical redshift error $\sigma_z$ and the systematic line-of-sight shift $\Delta r_{\parallel}$. The former is consistent across redshift bins at approximately $3\,\hMpc$, while the latter is positive in all bins and shows a mild decreasing trend with redshift.

The transverse proximity effect amplitude $\xi_0^{\rm TP}$ increases with redshift. However, this parameter is degenerate with the assumed mean free path value, $\lambda_{\rm UV} = 300\,\hMpc$, which is held fixed in this analysis. As a result, the observed trend cannot be uniquely attributed to an evolution of the proximity effect amplitude itself. A dedicated analysis in which both parameters are varied simultaneously would be required for a physical interpretation.

\subsection{Metals, HCD parameters, and noise}

The silicon metal biases show no clear evidence for significant redshift evolution. A mild trend is observed for $b_{\rm SiII(1193)}$, but it is not consistently supported by the other transitions and is only marginally significant; given parameter degeneracies, we do not attempt a physical interpretation. This caveat applies in particular to $b_{\rm SiIII(1207)}$: owing to its proximity to \lya, the associated excess peaks at $r_\parallel\simeq20$--$22\,\hMpc$, below the lower limit of our fitting range ($r>30\,\hMpc$), so it is less constrained than the other Si lines in all three bins. In the highest-redshift bin its posterior lies close to the $b\leq0$ prior boundary, with a strongly asymmetric uncertainty, and differs from the other two bins by at most $2\sigma$. Part of this behavior may arise from effects not captured by the baseline model at higher redshift, which we will investigate in future work.

The C\,\textsc{IV} bias and the HCD parameters $\beta_{\rm HCD}$ and $L_{\rm HCD}$ are largely prior-dominated and show no significant variation across redshift bins.

Finally, the correlated-noise amplitude $a_{\rm noise}$ decreases with redshift, as expected from the increasing DESI throughput at higher observed wavelengths, which improves the continuum signal-to-noise ratio and reduces the relative contribution of sky noise.


\section{Author Affiliations}
\label{sec:affiliations}

\noindent \hangindent=.5cm $^{a}${Institut d'Astrophysique de Paris, 98 bis boulevard Arago, 75014 Paris, France}

\noindent \hangindent=.5cm $^{b}${IRFU, CEA, Universit\'{e} Paris-Saclay, F-91191 Gif-sur-Yvette, France}

\noindent \hangindent=.5cm $^{c}${Lawrence Berkeley National Laboratory, 1 Cyclotron Road, Berkeley, CA 94720, USA}

\noindent \hangindent=.5cm $^{d}${Departamento de F\'{\i}sica, DCI-Campus Le\'{o}n, Universidad de Guanajuato, Loma del Bosque 103, Le\'{o}n, Guanajuato C.~P.~37150, M\'{e}xico}

\noindent \hangindent=.5cm $^{e}${Center for Astrophysics $|$ Harvard \& Smithsonian, 60 Garden Street, Cambridge, MA 02138, USA}

\noindent \hangindent=.5cm $^{f}${NASA Einstein Fellow}

\noindent \hangindent=.5cm $^{g}${Department of Physics, Boston University, 590 Commonwealth Avenue, Boston, MA 02215 USA}

\noindent \hangindent=.5cm $^{h}${Institute for Astronomy, University of Edinburgh, Royal Observatory, Blackford Hill, Edinburgh EH9 3HJ, UK}

\noindent \hangindent=.5cm $^{i}${Dipartimento di Fisica ``Aldo Pontremoli'', Universit\`a degli Studi di Milano, Via Celoria 16, I-20133 Milano, Italy}

\noindent \hangindent=.5cm $^{j}${INAF-Osservatorio Astronomico di Brera, Via Brera 28, 20122 Milano, Italy}

\noindent \hangindent=.5cm $^{k}${Department of Physics \& Astronomy, University College London, Gower Street, London, WC1E 6BT, UK}

\noindent \hangindent=.5cm $^{l}${Department of Physics and Astronomy, The University of Utah, 115 South 1400 East, Salt Lake City, UT 84112, USA}

\noindent \hangindent=.5cm $^{m}${Instituto de F\'{\i}sica, Universidad Nacional Aut\'{o}noma de M\'{e}xico,  Circuito de la Investigaci\'{o}n Cient\'{\i}fica, Ciudad Universitaria, Cd. de M\'{e}xico  C.~P.~04510,  M\'{e}xico}

\noindent \hangindent=.5cm $^{n}${NSF NOIRLab, 950 N. Cherry Ave., Tucson, AZ 85719, USA}

\noindent \hangindent=.5cm $^{o}${University of California, Berkeley, 110 Sproul Hall \#5800 Berkeley, CA 94720, USA}

\noindent \hangindent=.5cm $^{p}${Instituci\'{o} Catalana de Recerca i Estudis Avan\c{c}ats, Passeig de Llu\'{\i}s Companys, 23, 08010 Barcelona, Spain}

\noindent \hangindent=.5cm $^{q}${Institut de F\'{i}sica d’Altes Energies (IFAE), The Barcelona Institute of Science and Technology, Edifici Cn, Campus UAB, 08193, Bellaterra (Barcelona), Spain}

\noindent \hangindent=.5cm $^{r}${Departamento de F\'isica, Universidad de los Andes, Cra. 1 No. 18A-10, Edificio Ip, CP 111711, Bogot\'a, Colombia}

\noindent \hangindent=.5cm $^{s}${Observatorio Astron\'omico, Universidad de los Andes, Cra. 1 No. 18A-10, Edificio H, CP 111711 Bogot\'a, Colombia}

\noindent \hangindent=.5cm $^{t}${Institut d'Estudis Espacials de Catalunya (IEEC), c/ Esteve Terradas 1, Edifici RDIT, Campus PMT-UPC, 08860 Castelldefels, Spain}

\noindent \hangindent=.5cm $^{u}${Institute of Cosmology and Gravitation, University of Portsmouth, Dennis Sciama Building, Portsmouth, PO1 3FX, UK}

\noindent \hangindent=.5cm $^{v}${Institute of Space Sciences, ICE-CSIC, Campus UAB, Carrer de Can Magrans s/n, 08913 Bellaterra, Barcelona, Spain}

\noindent \hangindent=.5cm $^{w}${Fermi National Accelerator Laboratory, PO Box 500, Batavia, IL 60510, USA}

\noindent \hangindent=.5cm $^{x}${Department of Astronomy, University of Texas at Austin, 2515 Speedway, TX 78712, USA}

\noindent \hangindent=.5cm $^{y}${Center for Cosmology and AstroParticle Physics, The Ohio State University, 191 West Woodruff Avenue, Columbus, OH 43210, USA}

\noindent \hangindent=.5cm $^{z}${Department of Physics, The Ohio State University, 191 West Woodruff Avenue, Columbus, OH 43210, USA}

\noindent \hangindent=.5cm $^{aa}${The Ohio State University, Columbus, 43210 OH, USA}

\noindent \hangindent=.5cm $^{ab}${Department of Physics, University of Michigan, 450 Church Street, Ann Arbor, MI 48109, USA}

\noindent \hangindent=.5cm $^{ac}${University of Michigan, 500 S. State Street, Ann Arbor, MI 48109, USA}

\noindent \hangindent=.5cm $^{ad}${Department of Physics, The University of Texas at Dallas, 800 W. Campbell Rd., Richardson, TX 75080, USA}

\noindent \hangindent=.5cm $^{ae}${Department of Astronomy \& Astrophysics, University of Toronto, Toronto, ON M5S 3H4, Canada}

\noindent \hangindent=.5cm $^{af}${Department of Physics, Southern Methodist University, 3215 Daniel Avenue, Dallas, TX 75275, USA}

\noindent \hangindent=.5cm $^{ag}${Department of Physics and Astronomy, University of California, Irvine, 92697, USA}

\noindent \hangindent=.5cm $^{ah}${Sorbonne Universit\'{e}, CNRS/IN2P3, Laboratoire de Physique Nucl\'{e}aire et de Hautes Energies (LPNHE), FR-75005 Paris, France}

\noindent \hangindent=.5cm $^{ai}${Departament de F\'{i}sica, Serra H\'{u}nter, Universitat Aut\`{o}noma de Barcelona, 08193 Bellaterra (Barcelona), Spain}

\noindent \hangindent=.5cm $^{aj}${Department of Astronomy, The Ohio State University, 4055 McPherson Laboratory, 140 W 18th Avenue, Columbus, OH 43210, USA}

\noindent \hangindent=.5cm $^{ak}${Department of Physics and Astronomy, Siena University, 515 Loudon Road, Loudonville, NY 12211, USA}

\noindent \hangindent=.5cm $^{al}${Instituto Avanzado de Cosmolog\'{\i}a A.~C., San Marcos 11 - Atenas 202. Magdalena Contreras. Ciudad de M\'{e}xico C.~P.~10720, M\'{e}xico}

\noindent \hangindent=.5cm $^{am}${Instituto de Estudios Astrof\'isicos, Facultad de Ingenier\'ia y Ciencias, Universidad Diego Portales, Av. Ej\'ercito Libertador 441, Santiago, Chile}

\noindent \hangindent=.5cm $^{an}${Steward Observatory, University of Arizona, 933 N. Cherry Avenue, Tucson, AZ 85721, USA}

\noindent \hangindent=.5cm $^{ao}${Department of Physics and Astronomy, University of Waterloo, 200 University Ave W, Waterloo, ON N2L 3G1, Canada}

\noindent \hangindent=.5cm $^{ap}${Perimeter Institute for Theoretical Physics, 31 Caroline St. North, Waterloo, ON N2L 2Y5, Canada}

\noindent \hangindent=.5cm $^{aq}${Waterloo Centre for Astrophysics, University of Waterloo, 200 University Ave W, Waterloo, ON N2L 3G1, Canada}

\noindent \hangindent=.5cm $^{ar}${Space Sciences Laboratory, University of California, Berkeley, 7 Gauss Way, Berkeley, CA  94720, USA}

\noindent \hangindent=.5cm $^{as}${Instituto de Astrof\'{i}sica de Andaluc\'{i}a (CSIC), Glorieta de la Astronom\'{i}a, s/n, E-18008 Granada, Spain}

\noindent \hangindent=.5cm $^{at}${Departament de F\'isica, EEBE, Universitat Polit\`ecnica de Catalunya, c/Eduard Maristany 10, 08930 Barcelona, Spain}

\noindent \hangindent=.5cm $^{au}${Universit\'{e} Clermont-Auvergne, CNRS, LPCA, 63000 Clermont-Ferrand, France}

\noindent \hangindent=.5cm $^{av}${Department of Physics and Astronomy, Sejong University, 209 Neungdong-ro, Gwangjin-gu, Seoul 05006, Republic of Korea}

\noindent \hangindent=.5cm $^{aw}${Queensland University of Technology,  School of Chemistry \& Physics, George St, Brisbane 4001, Australia}

\noindent \hangindent=.5cm $^{ax}${Abastumani Astrophysical Observatory, Tbilisi, GE-0179, Georgia}

\noindent \hangindent=.5cm $^{ay}${Department of Physics, Kansas State University, 116 Cardwell Hall, Manhattan, KS 66506, USA}

\noindent \hangindent=.5cm $^{az}${CIEMAT, Avenida Complutense 40, E-28040 Madrid, Spain}

\noindent \hangindent=.5cm $^{ba}${Max Planck Institute for Extraterrestrial Physics, Gie\ss enbachstra\ss e 1, 85748 Garching, Germany}

\noindent \hangindent=.5cm $^{bb}${Department of Physics \& Astronomy, Ohio University, 139 University Terrace, Athens, OH 45701, USA}

\end{document}